\documentclass[twocolumn]{aastex631}
\usepackage{amsmath}
\usepackage{float}
\usepackage{soul}
\usepackage{graphicx}

\makeatletter
\newcommand*{\rom}[1]{\expandafter\@slowromancap\romannumeral #1@}
\makeatother

\begin{document}

\title{Uncovering deterministic behavior of black hole IGR J17091–3624: A twin of GRS 1915+105}

\author{Anindya Guria}
\affiliation{Department of Physics, Indian Institute of Science \\
C. V. Raman Road, \\
Bangalore 560012, India}
\thanks{anindyaguria@iisc.ac.in ; bm@iisc.ac.in}
\author{Banibrata Mukhopadhyay}
\affiliation{Department of Physics, Indian Institute of Science \\
C. V. Raman Road, \\
Bangalore 560012, India}

\begin{abstract}

Understanding nonlinear properties in accreting systems, particularly for black holes, from observation is illuminating as they are expected to be general relativistic magnetohydrodynamic flows that are nonlinear. Two features associated with nonlinear systems, used commonly, are chaos, which is deterministic, and random, which is stochastic. The differentiation between chaotic and stochastic systems is often considered to quantify the nonlinear properties of an astrophysical system. The particular emphasis is that data is often noise-contaminated and finite. We examine the dual nature of the black hole X-ray binary IGR J17091–3624, whose behavior has been closely studied in parallel to GRS 1915+105. Certain similarities in the temporal classes of these two objects are explored in literature. However, this has not been the case with their non-linear dynamics: GRS 1915+105 shows signs of determinism and stochasticity both, while IGR J17091-3624 was found to be predominantly stochastic. Here, we confront the inherent challenge of noise contamination faced by previous studies, particularly Poisson noise, which adversely impacts the reliability of non-linear results. We employ several denoising techniques to mitigate noise effects and employ methods like Principal Component Analysis, Singular Value Decomposition, and Correlation Integral to isolate the deterministic signatures. We have found signs of determinism in IGR J17091–3624, thus supporting the hypothesis of it being similar to GRS 1915+105, even as a dynamical system. Our findings not only shed light on the complex nature of IGR J17091-3624 but also pave the way for future research employing noise-reduction techniques to analyze non-linearity in observed dynamical systems.

\end{abstract}


\section{Introduction} \label{sec:intro}

Black hole accretion systems provide unique laboratories for studying nonlinear dynamics and general relativistic effects in extreme environments. The interplay of instabilities, turbulence and various accretion geometries in these systems manifests as complex temporal and spectral behaviors in our observation, making them invaluable for understanding fundamental accretion physics.

GRS 1915+105 stands as a particularly notable black hole X-ray binary, exhibiting high spin and a wide range of quasi-periodic oscillations (QPOs) \citep{Mirabel1994, belloni2001QPO}. This microquasar exhibits jets in certain temporal classes and was initially categorized into twelve temporal classes across three spectral states \citep{belloni2000}. Subsequent observations revealed two additional temporal classes: $\omega$ \citep{Klein-Wolt2002} and $\xi$ \citep{Hannikainen2003FirstIO}. All the features support the idea that the system is highly nonlinear when the strong general relativistic effects appear to control the source properties.

Early studies \citep{Mukhopadhyay2004, Misra_2004} demonstrated that some temporal classes of GRS 1915+105 display deterministic: chaotic (C), or fractal characteristics, while others appear stochastic (S). This classification is particularly significant given that accretion flows inherently involve instability and turbulence, which play a central role in matter transport and determining flow geometry. The correspondence between these nonlinear effects and the C or S nature of the flow provides crucial insights into accretion dynamics. The further question is: are the spectral characteristics (i.e., diskbb and power-law domination) also coupled to nonlinear features? \cite{Adegoke2018} advanced this understanding by combining timeseries behavior based on nonlinear features with spectral nature, identifying four distinct accretion regimes. It turns out for GRS 1915+105 that, for a given spectral feature, either kind of nonlinearity, C or S, is possible. This results in four combined spectral states and temporal classes. 
Earlier, one of us led arguing that \citep{Adegoke2018} (1) diskbb+C is the standard accretion disk \citep{Shakura-Sunyaev_1973}, (2) diskbb+S is the slim radiation trapped accretion disk \citep{Abramowics1988}, (3) power-law+C is the advection dominated accretion flow (ADAF: \citealt{Narayan&Yi1994}), and (4) power-law+S is the general advective accretion flow (GAAF: \citealt{Rajesh&Mukhopadhyay2010}).

IGR J17091-3624 is often considered GRS 1915+105's twin due to shared features like ``heartbeat" oscillations (see, e.g., \citealt{Janiuk2015} for possible physics), high-frequency QPOs, and different temporal classes \citep{Altamirano_2011,Rao_2012}. However, the former is fainter (see, e.g., \citealt{King2012, Miller2016}, who discuss the wind and its launching radii, which influence accretion budget and probably the source brightness) and possesses lower mass and spin \citep{Rao_2012}. Nevertheless, previous studies classified its temporal classes as S or noise-dominated \citep{Adegoke2020}, as opposed to the temporal classes of GRS 1915+105, which exhibit both C and S. This finding raises questions about the true nature of the dynamics of IGR J17091-3624, especially considering the significant contamination of X-ray lightcurves by Poisson noise. While past research suggested increasing bin size might influence fractal dimensions, conclusive results were elusive due to limited data \citep{Adegoke2020}. A more sophisticated approach is needed to preserve the underlying dynamics while handling Poisson noise.

We address this challenge through a comprehensive analysis of IGR J17091-3624's temporal classes using RXTE satellite data (from the HEASARC website\footnote{\url{https://heasarc.gsfc.nasa.gov/docs/archive.html}}) with a particular focus on addressing the noise issue. Our approach has two key aspects:

\begin{itemize}
    \item \textbf{Noise Reduction Techniques:} We will implement various filtering techniques to minimize the impact of Poisson noise on the lightcurves. This includes employing moving averages, non-local means filtering \citep{buades2004image, Buades_NLM2005}, and an adaptive algorithm optimized for detecting determinism in environments with significant noise \citep{Tung2011}.
    \item \textbf{Expanding Beyond Traditional Methods:} We will move beyond the correlation integral (CI) method, which has been most commonly used in this literature so far for S vs. C or even non-stochastic (NS) classification \citep{GRASSBERGER1983189, Misra:2006qg, harikrishnan2006, Adegoke2018, Adegoke2020}. This exploration will incorporate principal component analysis (PCA) \citep{PCA_SVD} and singular value decomposition (SVD) \citep{Misra:2006qg}, which have been rarely used so far for this purpose and hardly used to explore the consolidated conclusions. 
\end{itemize}

Exploring timeseries analysis of black hole sources based on various filtering techniques, as outlined above, is a novel initiative, as of this work. Through the present comprehensive analysis, we expect to find evidence suggesting several of IGR J17091-3624's temporal classes exhibit at least NS behavior, if not C, potentially refuting the earlier findings based on unfiltered data \citep{Adegoke2020}. We also combine our timeseries analysis results with the spectral analysis of \cite{Adegoke2020} to draw a clearer picture of IGR J17091-3624's accretion states. This work highlights the crucial role of noise reduction techniques and the development of innovative testing methods for effective non-linear analysis of noisy astrophysical timeseries data. Each of our methods is based on some heuristic expectations about the general nature of chaotic timeseries and Poisson noise.

Although handling Poisson noise from the power spectrum (e.g., for QPO identification) is well-known \citep{Uttley2014}, denoising in the time domain comes with additional challenges. This is because
the presence of noise not only affects the Fourier amplitudes but also adds a random component to the phases, which is unimportant for the power spectrum. However, this information is
essential to reconstruct the dynamics of the system (our present goal). We expect the methods presented by us to be useful in other applications too, where the timing properties are necessary to be extracted amidst Poisson noise.

The next section will discuss the specific challenges encountered during the non-linear analysis of IGR J17091-3624 lightcurves and the solutions implemented. We will then provide detailed descriptions of each filtering algorithm and the classification methods used to differentiate between S and NS classes for both the filtered and unfiltered lightcurves in sections \ref{sec:proposed filtering} and \ref{sec:Classification}, respectively. Subsequently, we will present the results for all combinations of filtering and testing methods considered in section \ref{sec:results}. As with previous endeavors \citep{Adegoke2018, Adegoke2020}, we aim to investigate further the underlying accretion processes responsible for the observed non-linear features. Finally, we end with a summary and conclusion in \ref{sec:Discussion}.

Throughout this work, we have often used the abbreviation ``IGR" in place of IGR~J17091–3624 for brevity. The individual temporal classes of IGR are labeled using Roman numerals from \rom{1} to \rom{9}. Hence, e.g., by IGR-\rom{5} we refer to the fifth temporal class of IGR~J17091–362. Also, occasionally, we refer to GRS~1915+105 as GRS.

\section{Difficulties encountered earlier with IGR~J17091-3624} \label{sec:difficulties}
Several caveats exist in determining whether an underlying dynamical system is S or NS. It is challenging in an experimental or observational context, where the data is of limited length and almost inevitably contaminated by noise.

In the case of IGR~J17091-3624, previous studies found no signs of NS nature. Simultaneously, it was also noted that compared to a source like GRS 1915+105 (whose deterministic nature has been well-established, at least for some temporal classes), the average photon count recorded by \textit{RXTE/PCA} is about $\sim20$ times less. This leads to a very high ratio of expected Poisson noise $\langle PN\rangle$ to the RMS variation of photon counts. Poisson noise arises from the characteristics of photon counting devices, which conform to a Poisson distribution. Table \ref{tab:Poisson Stat} summarizes the expected Poisson noise in each of the temporal classes of IGR~J17091-3624. This noise is primarily attributed to the discrete nature of light particles and the independent nature of photon detections. As the number of photons decreases, the influence of Poisson noise increases proportionally. Hence, conclusions regarding non-linear properties from such noise-contaminated data must be taken cautiously.

There have been hints by \cite{Adegoke2020} towards deviation from stochasticity when the binning of the timeseries data had been increased from $0.125s$ to $0.5s$ or higher. However, definitive conclusions could not be drawn due to the significant reduction in the number of data points resulting from increased binning, adversely impacting the performance of the CI method they utilize.

Nevertheless, an increased binning effectively takes a mean of the incoming photon flux over a longer time scale, thus eliminating (nonetheless crudely) some of the Poisson noise effects. If this is true, this observation indicates that with more sophisticated noise-removal techniques, one could analyze this peculiar conundrum of GRS and IGR from a non-linear perspective.

\begin{deluxetable*}{cccccccc}
\colnumbers
\tabletypesize{\scriptsize}
\tablewidth{0pt} 
\tablecaption{Poisson noise statistics of IGR J17091-3624\label{tab:Poisson Stat}}

\tablehead{
\colhead{ObsID} & \colhead{Class}& \colhead{GRS-like class}& \colhead{Data points} & \colhead{$\langle S \rangle$} & \colhead{RMS} &\colhead{$\langle PN \rangle = \langle S \rangle^{1/2}$} &\colhead{$\langle PN \rangle$/rms}\\
}
\startdata
{96420-01-01-00} & {\rom{1}} & {$\chi$}    & {13424} & {13.78} & {4.70}  & {3.71} & {0.79} \\
{96420-01-11-00} & {\rom{2}} & {$\phi$}    & {26713} & {15.95} & {5.29}  & {3.99} & {0.76} \\
{96420-01-04-01} & {\rom{3}} & {$\nu$}     & {9217}  & {22.74} & {6.51}  & {4.77} & {0.73} \\
{96420-01-05-00} & {\rom{4}} & {$\rho$}    & {21697} & {29.50} & {9.47}  & {5.43} & {0.57} \\
{96420-01-06-03} & {\rom{5}} & {$\mu$}     & {14561} & {12.38} & {7.55}  & {3.52} & {0.47} \\
{96420-01-09-00} & {\rom{6}} & {$\lambda$} & {9953}  & {30.84} & {9.79}  & {5.55} & {0.57} \\
{96420-01-18-05} & {\rom{7}} & {None}      & {5169}  & {26.78} & {16.33} & {5.18} & {0.32} \\
{96420-01-19-03} & {\rom{8}} & {None}      & {13025} & {30.83} & {16.55} & {5.55} & {0.34} \\
{96420-01-35-02} & {\rom{9}} & {$\gamma$}  & {9240}  & {28.28} & {9.61}  & {5.32} & {0.55} \\
\enddata

\tablecomments{The columns show the following: (1) \textit{RXTE} ObsID from where the data has been taken, (2) temporal class based on classification by \cite{Altamirano_2011}, (3) corresponding GRS-like class, (4) number of data points in the time-series for the mentioned ObsID, (5) average photon count in lightcurve $\langle S \rangle$, (6) the root mean squared (rms) variation in photon count, (7) expected variation due to Poisson noise, $\langle PN \rangle = \langle S \rangle^{1/2}$, and (8) the ratio of expected Poisson noise to the actual rms variation. The Poisson noise appears to be a significant contribution to the rms in all temporal classes.}

\end{deluxetable*}

\section{Proposed filtering methods} \label{sec:proposed filtering}
Based on the experience from previous efforts, we can list some key properties that any filtering method used for our application should ideally satisfy. The rationale behind many of these points has been discussed succinctly by \cite{Kostelich1993}.

\begin{enumerate}
    \item There should be no implicit or explicit dependence on the signal's power spectrum. This is because chaotic timeseries are expected to have a broad spectral spread. Also, the power spectrum of Poisson noise of an ideal Poisson process is uniform: traditional low pass or band pass filtering would not be effective in our application.
    \item Our filtering technique should avoid introducing additional local correlation in our data. It would ensure that our conclusions are not affected by the filtering process.
    \item The number of data points should not be reduced drastically by the filtering algorithm. As already talked about, this enables the use of the CI method. In this work, we utilize the time resolution of 0.125 s from RXTE observations, which appears to be an optimum choice for the present purpose.
   
\end{enumerate}
We first describe some trivial filters such as boxcar and Gaussian convolution, which, although imperfect, improve upon the previous attempt of increasing the binning of the lightcurve. We then discuss some more sophisticated tools like non-local means denoising and adaptive denoising, which incorporate more of the criteria listed above. Hence, we proceed in implementing denoising from simpler to more advanced methods. In section \ref{subsec: comparison denoise}, we have compared their effects on IGR lightcurves.

\subsection{Convolution-based filters}

Here, we briefly review convolution kernels. Detailed discussions can be found in any standard digital signal processing text (e.g., \citealt{Smith_Steven_W_1997}). Let us use a kernel $g_i$ of length $2k+1$ to filter the original (input) timeseries $x_n$. We are considering only normalized kernels $\left(\sum_{i=-k}^{k}g_i=1\right)$ for the sake of brevity. The output timeseries $y_n$ is defined as the convolution:
\begin{equation}
    y_n=\sum_{i=-k}^{k}g_ix_{n+i}.
\end{equation}
The free parameters are the kernel's nature and length. We have used the two most popular kernels for filtering applications, boxcar and Gaussian. In our study, we do not employ padding at the start and end of the series, as it might lead to spurious results. Although this implies reducing the signal's length by $2k$, we deem this trade-off acceptable to retain only the valid data points and no padded values.
\subsubsection{Moving average filter (BOX)}
The moving average filter (also called a boxcar filter due to its kernel's shape) is commonly used in digital signal processing. 

This technique convolves the input signal with an equal-weight kernel. They are the most efficient in reducing white noise while preserving the sharpness of step response. It is a popular example of convolution in the context of time-domain encoded signals. The kernel and filtered data are defined as:
\begin{align}
    g_i&=\frac{1}{2k+1},\\
    y_n&=\frac{1}{2k+1} \sum_{i=n - k}^{n + k}x_{i}.
\end{align}

We have chosen this method since it closely resembles the effect of binning. In section \ref{sec:difficulties}, we have discussed how increased binning helps alleviate some of the noise contamination at the expense of the number of data points. With binning, one essentially takes an ``average" of the signal over some time period. Mathematically, binning a series $\{x_n\}$ of length $N$ into bins of size $2k+1$ is:
\begin{equation}    
y_m = \frac{1}{2k + 1} \left( \sum_{n = (m-1)(2k + 1)}^{m(2k + 1)} x_n \right),
\end{equation}
where $y_m$ is the $m^{th}$ element of the binned series. The resulting series $\{y_k\}$ is of length $\simeq N/(2k + 1)$. Compared to this, the moving average has the benefit of having $\simeq N - 2k$ data points, which meets one of our three criteria.

We tried to use the shortest kernel length possible to minimize the introduction of local correlations while still improving the Signal-Noise Ratio (SNR). We have chosen $2k+1= 9$ in our presented results. Refer to Table \ref{tab:SNR_sample} for quantitative details.

There are quite a few drawbacks to the boxcar convolution method. First, it introduces local correlation to the signal which is undesirable. Second, its effect on the signal's power spectra is known to be that of a low-pass filter (sinc function). Additionally, due to the discontinuous nature of this kernel, its frequency response has unwanted ripples. We move on to more involved methods that address these issues.
explicitly preserved during boxcar denoising.

\subsubsection{Gaussian filter (GAU)}
\label{subsec: gaussian filter}

From the central limit theorem, we know that an infinite number of passes by a boxcar filter tends to convolution by a Gaussian kernel. Since the Fourier transform of a Gaussian is also a Gaussian distribution, the frequency response of this kernel has no ripple effect. Hence, it is an improvement over our previous moving averages approach.

The free parameters for Gaussian filtering are the spread $\sigma$ and the kernel length. The formulas for the kernel and the convoluted signal are
\begin{equation}
\begin{aligned}
g_i&=Ce^{-\frac{i^2}{2\sigma^2}}\\
y_n&=C\sum_{i=-k}^{k}e^{-\frac{i^2}{2\sigma^2}}x_{n+i}
\end{aligned}
\end{equation}
respectively. Here, $C$ is a normalization constant. Note that this $\sigma$ is not related to the actual standard deviation of the signal; rather, it quantifies how many of the data points in the vicinity of available data $x_n$ effectively contribute to the denoised data $y_n$. We have chosen $\sigma=3$, with a kernel length of ${  9}$. The kernel length does not matter much for Gaussian filtering since the edge values quickly fall off to $\sim 0$.

Although an improvement from boxcar convolution, this method still possesses many of the same deficiencies. It is still essentially a low-pass filter in the frequency domain, and it also introduces local correlations. To overcome these issues, we adopt a drastically different approach, as outlined in the following subsections.

\subsection{Adaptive Denoising Algorithm (ADA)}
Proposed by \cite{Tung2011}, this technique is suitable for detecting chaos/determinism within heavy noise. Their work showed that it performs better than wavelet shrinkage and other chaos-based approaches. Here, we briefly describe ADA and its implementation.

Initially, the algorithm divides a timeseries into segments or windows of length $w = 2n + 1$ points, with $n+1$ points overlapping among them. A $k^{th}$ order polynomial is fitted for each segment. For the $i^{th}$ and $(i+1)^{th}$ segments let them be $p^k_i(l_1)$ and $p^k_{i+1}(l_2)$, where $l_1,l_2\in\{1,2,\cdots 2n+1\}$. The denoised data within this $i^{th}$ overlapped region are found using the weights $w_1=\left[1-(l-1)/n\right]$ and $w_2=\left[(l-1)/n\right]$ as
\begin{equation}
\begin{array}{cc}
y_{ni+l}=w_1 p^k_i(l+n)+w_2 p^k_{i+1}(l) & l\in\{1,2,\cdots n+1\}
\end{array}.
\end{equation}
The weights decrease linearly with the distance between the segments' point and center. This choice of weighting eliminates discontinuities at the boundaries of neighboring segments; this fitting ensures continuity all across the data.

Two free parameters for fine-tuning are the segment size $(w)$ and polynomial order $(k)$. It shall be observed that the original signal can be reconstructed back for sufficiently small segments and a high order of polynomial fitting.

Based on the original paper by \cite{Tung2011}, we have followed the recommended protocol to determine the optimum values. Here, we have used $5^{th}$ order polynomial fit to maximize the closeness of the polynomial fit to the real data. While varying the segment size for a fixed $k=5$ (our choice of order), the RMS error between the denoised series and actual data has been measured. For (i) small $w$, this error first increases, (ii) then nearly flattens out for some range of $w$, and finally, (iii) for even larger segment size, there is a sharp increase. (i) corresponds to nearly perfect fitting with almost no denoising, while (iii) means all local variations have been suppressed. Therefore, we choose $w$ for each lightcurve to correspond to the least slope of the deviation plot (corresponding to ii). For most cases, we find $w\sim15-20$. Although changing $k$ beyond $3$ does not lead to any drastic difference in RMS error, we choose $5$ since the additional computation time is acceptable. 

Indeed ADA is an improvement over the convolution-based filters, since this explicitly tries to model the non-linearity of the signals (with polynomial regression). However, this is still not ideal, as its frequency response does introduce artifacts in the frequency domain. Section \ref{subsec: comparison denoise} demonstrates a few examples of this effect.

\subsection{Non-local means (NLM) denoising}
Initially devised for image denoising, the non-local means (NLM) algorithm mitigates noise by leveraging similar samples or pixels, regardless of their spatial proximity. This non-local approach does not require additional assumptions imposed on the data other than the presence of redundant structures within it.

We have suitably adjusted the algorithm for denoising a one-dimensional signal and laid out a way to robustly choose the necessary parameters. The assumption of redundant structures in our data is justified, as truly chaotic timeseries typically should exhibit underlying self-similarity across multiple time scales. We review our implementation of this algorithm here; a detailed discussion can be found in \citealt{buades2004image, Buades_NLM2005}.

In this method, we first construct a list of vectors $\{\vec{v}_i\} = \{(x_{i-k}, x_{i-k+1}, \cdots x_{i+k})\} $ of dimension $2k+1$ centered around each $x_i$. These are essentially $\sim N$ patches of size $2k+1$, among whom we are interested in finding the similarities and assigning the weight by which they influence the denoised timeseries. For that purpose, we define a two-point weight function denoted by the matrix
\begin{equation}
    f_{m,n} = \exp\left[{-\frac{{ \lVert \vec{v}_m-\vec{v}_n \rVert^{2}}}{{h_n}^{2}}}\right].
\end{equation}
The decay parameter $h_n$ quantifies how strongly the weight $f_{m,n}$ is affected by the deviation of the patch $\vec{v}_m$ from $\vec{v}_n $. The deviation itself is measured by the standard Euclidean norm $\Vert \cdot \Vert^2$. The resultant NLM denoised timeseries $\{y_n\}$ is given by,
\begin{equation}
    \begin{aligned}
y_n&=\frac{1}{C_n}\sum _{m=1}^{N}f_{m,n} x_m, \\
C_n&=\sum _{m=1}^{N}f_{m,n}.
\end{aligned}
\end{equation} 

\begin{figure*}[t!]
    \centering
    \includegraphics[width=\linewidth]{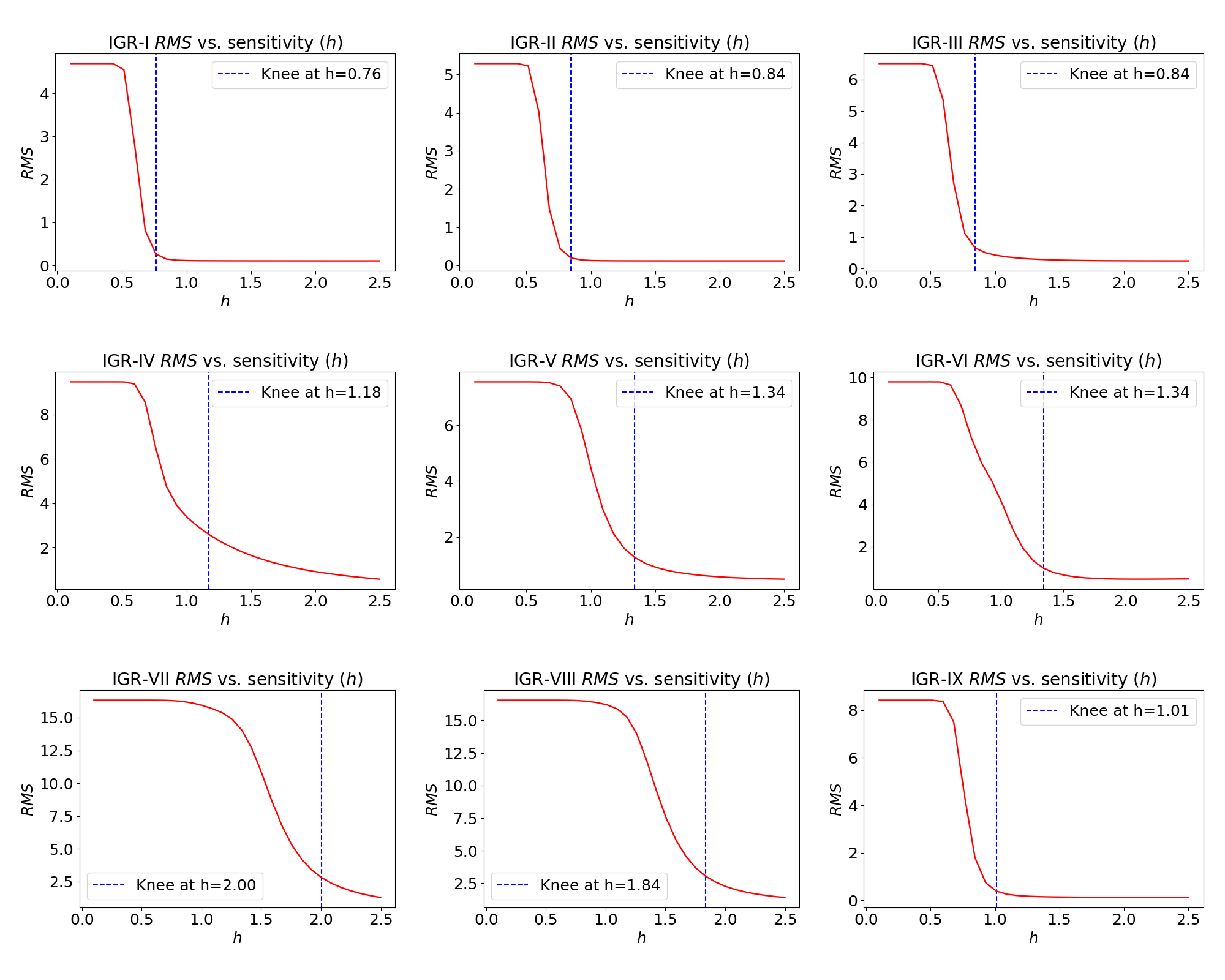}
    \caption{The variation of RMS with increasing sensitivity proportionality constant. This helps to choose the optimal sensitivity $h$ for NLM denoising of IGR lightcurves by finding the lower ``knee". }
    \label{fig: NLM sensitivity}
\end{figure*}
The choice of $h_n$ is critical to this filter's performance. It has been suggested earlier that $h_n$ should be of similar order to that of noise standard deviation (distinct from the signal's RMS, \citealt{Ville2009}). In earlier applications of NLM to denoise one-dimensional data (see, e.g., \citealt{NLM_ECG} with ECG signal), the noise standard deviation was known \textit{a priori} by comparison with simulated data as a dummy for ground truth. In a more traditional setting of using NLM in image processing, techniques have been suggested to estimate noise RMS
by manually selecting image patches with a uniform background and estimating the noise level by calculating the mean absolute deviation within that selection of patches (see, e.g., \citealt{Chen2015}). In our situation, we can not identify specific sections of our data and assume that they represent a constant background, as that can lead to spurious conclusions about S/NS classification. We do not have simulated lightcurves to compare with the observed lightcurve and estimate noise standard deviation.

We have already seen in Table \ref{tab:Poisson Stat} that the dominant source of noise in our available lightcurves is Poisson. Therefore, for the present purpose, we argue that it is logical to set $h_n$ to be proportional to the expected Poisson noise variance at the patch $\vec{v}_n$, i.e., the square root of the mean photon flux over this patch, given by
\begin{equation}
h_n \propto \sqrt{\langle  \vec{v}_n \rangle} = h \sqrt{\langle  \vec{v}_n \rangle}.
\end{equation}

For a given lightcurve, we need a way to set the proportionality constant $h$ robustly. We notice that upon increasing this sensitivity factor $h$, there is a steep decline in the RMS of the denoised lightcurve, giving an ``S"-curve, see, e.g., Figure \ref{fig: NLM sensitivity}. By finding $h$ corresponding to the lower ``knee" of this curve, we can fix this parameter without making any other assumption about the data.

A similar example of setting the sensitivity of NLM according to the expected Poisson noise level can be found in \citealt{NLM_Poisson2016}, where this procedure has been described as a weight correction of the decay parameter due to Poisson noise. This scheme is non-local since no implicit or explicit priority is given to the patches based on their proximity. Although a time scale is implicitly chosen when we set the size of the patches to assign the weights, this does not introduce any priority to the points within these patches in how they contribute to calculating the weights $f_{m,n}$. This assures that we are not introducing any additional local correlation through the denoising process. The NLM algorithm also leaves behind ``method noises". Previous studies showed this to be low amplitude white noise \citep{buades2004image}, which should not cause drastic problems in non-linear analysis. This method satisfies all the criteria that we had set out for an appropriate denoising algorithm.

\begin{figure*}
    \centering
    \includegraphics[width=\linewidth]{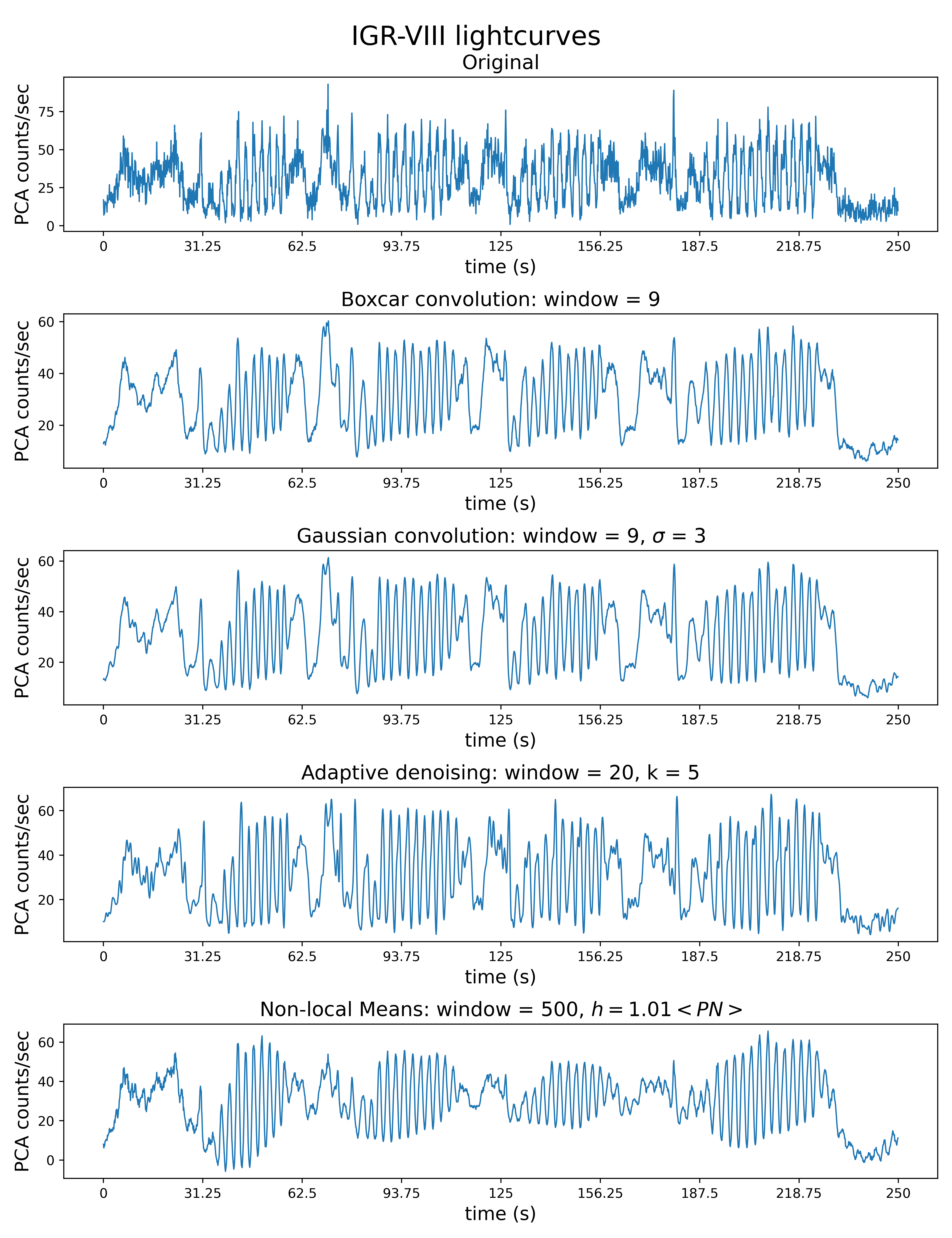}
    \caption{Pre and post filtering lightcurves of IGR-\rom{8}.}
    \label{fig: IGR-VIII timeseries}
\end{figure*}

\begin{figure*}
    \centering
    \includegraphics[width=0.9\linewidth]{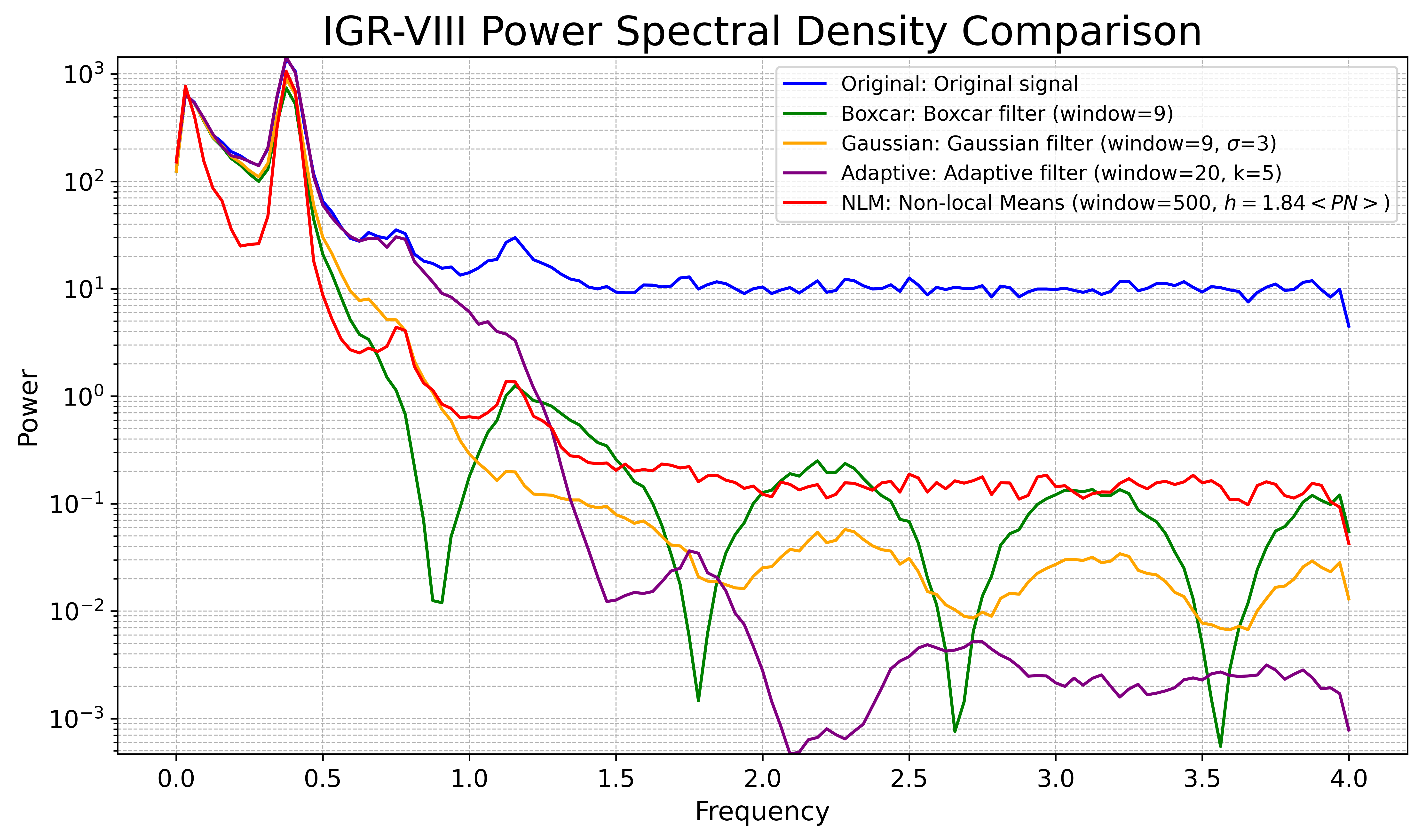}
    \caption{Power spectral density (PSD) of IGR-\rom{8} pre and post denoising.}
    \label{fig: IGR-VIII PSD}
\end{figure*}

\begin{figure*}
    \centering
    \includegraphics[width=0.9\linewidth]{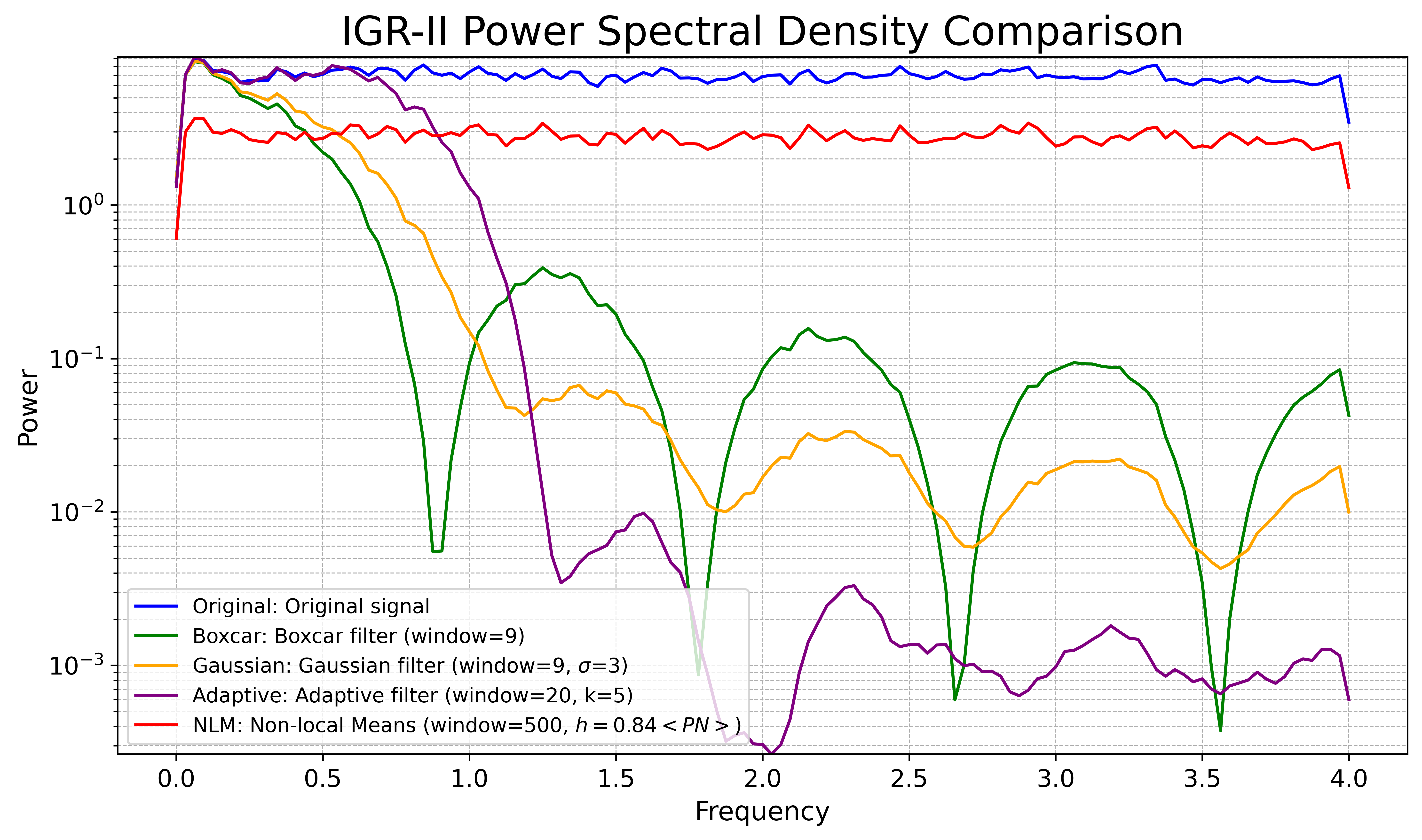}
    \caption{Power spectral density (PSD) of IGR-\rom{2} pre and post denoising.}
    \label{fig: IGR-II PSD}
\end{figure*}

\begin{figure*}
    \centering
    \includegraphics[width=\linewidth]{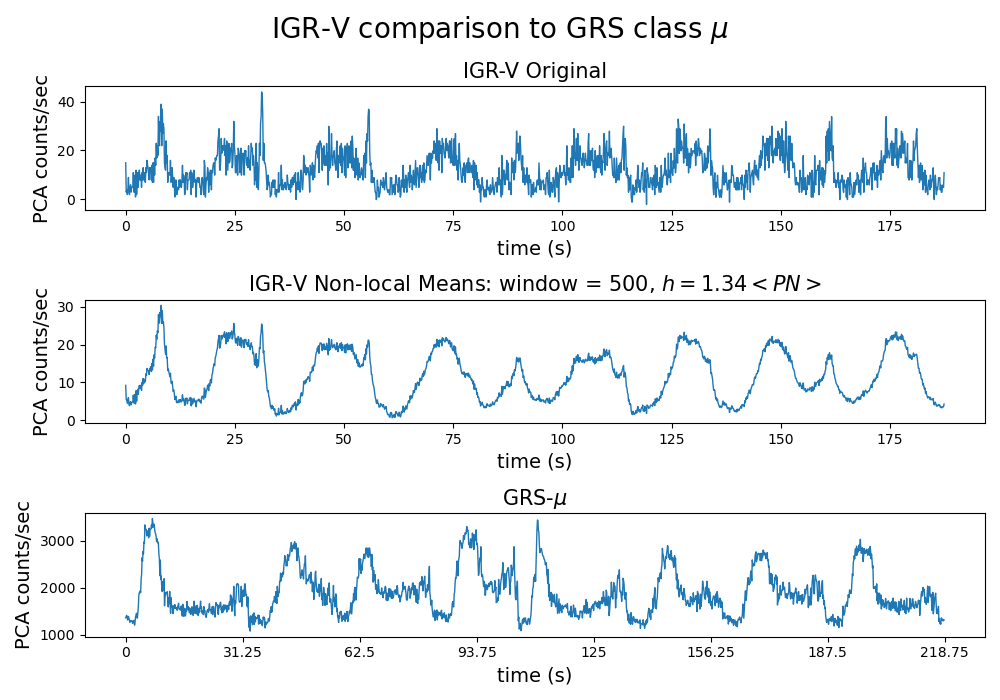}
    \caption{Comparison of IGR-\rom{5} lightcurves (original and NLM) with its GRS-1915+105 counterpart $\mu$.}
    \label{fig: IGR-V GRS-mu timeseries}
\end{figure*}

\begin{deluxetable}{ccccccc}
\tabletypesize{\scriptsize}
\tablewidth{\textwidth} 
\tablecaption{SNR estimation\label{tab:SNR_sample}}

\tablehead{
\colhead{Class}& \colhead{GRS-like class}& \colhead{original}& \colhead{BOX}& \colhead{GAU}& \colhead{ADA}& \colhead{NLM}
}
\colnumbers

\startdata 
\rom{1} & $\chi$& 10.19 & 19.14 & 18.21 & 15.36 & 9.93\\
\rom{2} & $\phi$& 10.15 & 19.33 & 19.12 & 17.86 & 7.29\\
\rom{3} & $\nu$& 12.24 & 21.14 & 20.57 & 18.04 & 17.96\\
\rom{4} & $\rho$& 14.11 & 21.51 & 21.35 & 19.18 & 20.70\\
\rom{5} & $\mu$& 6.37 & 8.88 & 8.71 & 8.11 & 8.70\\
\rom{6} & $\lambda$& 3.70 & 4.72 & 4.67 & 4.48 & 5.39\\
\rom{7} & None& 4.64 & 6.74 & 6.37 & 5.32 & 6.44\\
\rom{8} & None& 7.30 & 10.09 & 9.58 & 8.08 & 12.22\\
\rom{9} & $\gamma$& 10.07 & 15.71 & 15.28 & 13.70 & 14.06\\
\enddata

\tablecomments{The columns show the following: (1) temporal class based on \cite{Altamirano_2011}; (2) GRS-like class; (3), (4), (5), (6), and (7) are the SNR in $dB$ for original, BOX, GAU, ADA, and NLM respectively.}
\end{deluxetable}

\subsection{Comparison among the denoising techniques}
\label{subsec: comparison denoise}
Figure \ref{fig: IGR-VIII timeseries} illustrates the effect of all our filtering methods on IGR class \rom{8} (a class we later find to be NS; see section \ref{sec:Discussion}). The corresponding power spectral density (PSD) in Figure \ref{fig: IGR-VIII PSD} shows the effect of each filter in the frequency domain. The flat PSD at higher frequencies in Figure \ref{fig: IGR-VIII PSD}, and at all frequencies in Figure \ref{fig: IGR-II PSD}, is characteristic of Poisson white noise (as seen earlier in different related contexts in \citealt{Uttley2014}). The BOX and GAU methods seem to act as low-pass filters, as expected. The sinc-function frequency response of BOX leaves behind characteristic artifacts in the PSD. Since our BOX and GAU kernels are of size 9 (therefore, of period $= 9 \times 0.125s = 1.125s$ or frequency $\simeq 0.9 Hz$) we see dips at $0.9Hz$ and its higher harmonics. These artifacts are much less apparent for GAU (due to reasons explained in \ref{subsec: gaussian filter}). The ADA filter also effectively acts as a low-pass filter. This is expected since ADA works by fitting polynomials to windows of finite size in an attempt to capture the non-linearity in the signal. The spectral signature of NLM is interesting since the features in the PSD of the original signal at higher frequencies are only scaled down, while the lower frequency dynamics are preserved as it is. Since ideal Poisson noise has a uniform PSD, NLM's effect in the frequency domain is, therefore, promising for mitigating this issue. 
This is particularly significant for IGR \rom{2} in Figure \ref{fig: IGR-II PSD} (a S class as per this study). The original signal has a uniform PSD with no other features, which is retained after NLM. The other local method's frequency response is evident from their corresponding PSDs.

Further, in Figure \ref{fig: IGR-V GRS-mu timeseries}, we have shown a comparison between lightcurves of IGR-\rom{5} (a NS class as found by us) and its corresponding GRS-1915+105 counterpart $\mu$. Qualitatively, post denoising the similarities between the IGR and GRS classes are more apparent.

Since we do not have a reference model for the expected signal, we attempt at a naive approach to estimate the SNR of the denoised signals. The results have been reported in Table \ref{tab:SNR_sample}. To calculate these values, we have chosen a uniform patch of $\approx 12s$ (i.e. 100 data points) from the timeseries and estimated SNR as $20\log_{10}(<S>/RMS)$. The relative gain of the estimated SNR over this patch after denoising is more relevant to our discussion rather than the actual numerical value of the SNR itself, since the value can vary from one choice of patch to another. Evidently, BOX and GAU consistently increases the SNR by 1.5-2 times. ADA gains are comparable to them, although slightly less. NLM SNR gain is more interesting. For the classes which we later find to be S, NLM has lesser SNR improvement compared to the other methods (e.g. \rom{1}, \rom{2}, and \rom{3}). On the other hand, NLM seems to more effective for classes which we later find to be NS. This behavior is expected, since the working principle of NLM it to use redundant and repeated patterns in the signal for denoising, which is ineffective in the case of S signals. This property works in our favor as it ensures that the NLM method will not ``over-smoothen" an S signal, which is a potential pitfall of other filters. 

We would like to highlight that this method of calculating the SNR is not perfect, and ideally some model of the signal is essential for a robust estimation.

\section{Methods for Classification of S and NS Timeseries} \label{sec:Classification}
The researchers, particularly in this field of study, have predominantly relied on the CI method to understand the non-linearity in the lightcurve of a black hole source. However, recent advancements have introduced a range of alternative approaches that have shown comparable efficacy in achieving similar results. In light of these developments, we have incorporated some notable alternatives that have been demonstrated to be successful in classifying GRS~1915+105 temporal classes. We have made noteworthy improvements in overcoming practical challenges associated with utilizing some of these alternative techniques.

In this section, we first very briefly review the standard Grassberger-Procaccia (GP, \citealt{GRASSBERGER1983189}) CI algorithm. Following this, we introduce other methods.
\subsection{Correlation Integral Method} \label{subsec:CI method}
The delay-embedding technique of GP is used to reconstruct the dynamics of a non-linear system from a one-dimensional timeseries. Assuming that we are embedding in a reconstructed phase space of $M$ dimensions with a delay of $\tau$, the $i^{th}$ vector is constructed as
\[\Vec{\xi_i}=[x_{i},x_{i+\tau},x_{i+2\tau}\cdots x_{i+(M-1)\tau}],\]
where $x_i$ is the $i^{th}$ element of a timeseries. An implicit condition to be satisfied is that the timeseries must be temporally equally spaced. We have used a binning of $0.125 s$ for  IGR J17091–362 lightcurves, similar to \cite{Adegoke2020}. Along with this, the choice of delay needs to ensure no correlation among the components of these vectors. In literature, there are several available suggestions; we have chosen the one by \cite{Fraser1986} where $\tau$ is the first local minimum of the mutual-information function of $x_n$ or when it crosses $1/e$, whichever is shorter.

In this method, one finds the probability of two arbitrary vectors to be closer than a distance $r$, which is related to the correlation integral. In the case of a $M$-dimensional embedded data, it is denoted by $C_M(r)$, given by
\begin{equation}C_M(r)=\frac{1}{N_v(N_c-1)} \mathop{\sum^{N_v}\sum^{N_c}}_{i=1\  j\neq i\ j=1} H\left(r-\left\vert \Vec{\xi_i}-\Vec{\xi_j} \right\vert\right).\end{equation}
Here $N_v$ is the total number of vectors $\Vec{\xi_i}$, while $N_c$ is the number of vectors chosen randomly to estimate the probability. $H$ is the Heaviside step function. \cite{GRASSBERGER1983189} defined an estimate of the Hausdorff fractal dimension called the correlation dimension ($D_2$) as
\begin{equation}D_2=\lim_{r\to0}\left(\frac{d\log{C_M(r)}}{d\log{r}}\right).\end{equation}
The $D_2$ value gives the effective number of differential equations that govern the system's dynamics. Once $D_2$ is evaluated for several $M$, we can obtain some of the non-linear dynamical properties of the system. If for all $M$, $D_2\approx M$, then the system is S; this indicates no significant underlying structure in the phase space representation of the data. Otherwise, it is possibly deterministic if first $D_2$ increases linearly with increasing $M$ and then saturates to a constant value. There are additional criteria for signal contaminated with colored noise (which can also cause saturation of the $D_2$ value). We shall discuss this issue in the next section. The value of $M$, where saturation occurs, is the optimal embedding dimension, which we denote by $M_c$.
\subsubsection{Surrogate analysis} \label{subsubsec:Surrogate method}
Saturation of $D_2$ is a necessary but not sufficient condition for NS characteristics. Due to the possibility of contamination by colored/pink noise (noise featuring a $1/f^\alpha$ power spectrum), as seen in \cite{Misra:2006qg}, signals can also show saturation. Surrogate data analysis is commonly used to increase confidence in identifying non-trivial structures in timeseries data. It effectively distinguishes intrinsic fractal patterns from colored noise. The method involves formulating a null hypothesis that assumes the data to originate from a stationary linear stochastic process. By comparing the $D_2$ vs. $M$ plots of the surrogate timeseries with that of the true data, we check if the null hypothesis can be rejected (with $95\%$ confidence in our case). 

Surrogate data sets are generated to match the original data in its probability distribution and power spectrum. The iterative amplitude-adjusted Fourier transform (IAAFT) method is commonly employed for generating such surrogates. {  The IAAFT method begins by randomly shuffling the original time series to create an initial surrogate; this preserves the distribution of amplitudes at the expense of power spectrum. To fix that, this surrogate is then Fourier transformed, and its phases are randomized while adjusting the Fourier amplitudes to that of the original series. An inverse Fourier transform is then performed, followed by an adjustment of the amplitudes to match the original distribution. These steps are iteratively repeated, refining both the spectral properties and amplitude distribution until the surrogate closely matches the original time series in terms of both these characteristics. This ensures that the resulting surrogate data are suitable for rigorous hypothesis testing while retaining essential statistical properties of the original signal.} For a more comprehensive explanation of this method, refer to \cite{Venema_IAAFT2006} and related literature. In our study, we use the {\tt\string NoLiTSA} python package. The software is available on GitHub\footnote{\texttt{NoLiTSA} codebase: \url{https://github.com/manu-mannattil/nolitsa}.} under a 3-clause BSD license.

\cite{harikrishnan2006} proposed a statistic called the Normalised Mean Sigma Deviation (NMSD), which has been used earlier to quantify the differences between the values of $D_2$ of the surrogates and the signal (the discriminating factor). This measure is defined as
\begin{equation}n m s d^2=\frac{1}{M_{\max }-1} \sum_{M=M_{\min }}^{M_{\max }}\left[\frac{D_2(M)-\left\langle D_2^{\mathrm{surr}}(M)\right\rangle}{\sigma_{\mathrm{SD}}^{\mathrm{sur}}(M)}\right]^2,\end{equation}
where $M_{max}$ and $M_{min}$ are the minimum and maximum embedding dimensions, respectively, for the analysis. 
$\left\langle D_2^{\mathrm{surr}}(M)\right\rangle$ is the mean of $D_2$ values of the surrogates at a given $M$ and $\sigma_{\mathrm{SD}}^{\mathrm{sur}}(M)$ is the standard deviation. \cite{harikrishnan2006} showed that for 19 surrogates, a $nmsd < 3$ indicates noise domination in the data, and the null hypothesis cannot be rejected. However, when $nmsd > 3$, a distinction can be made between the surrogates and the true $D_2-M$ curves -- indicating NS processes in our timeseries. 

Therefore, to test for the NS property of a lightcurve, we generate 19 surrogates from it and obtain the $D_2-M$ curves for each of them (i.e., 20 total series). Upon calculating the $nmsd$ statistics from these $D_2$ data, if the $nmsd>3$ criterion is satisfied, we conclude the lightcurve to be NS. This process is carried out for all the original and denoised lightcurves of IGR.

\subsection{Singular Value Decomposition (SVD)}\label{subsec:SVD}
An interesting historical introduction and review of SVD can be found in \citealt{SVD_history}. \cite{BROOMHEAD1986217} provide rigorous mathematical details and motivation on the application of SVD for non-linear timesereis analysis on observed or experimental data. This method has been used previously for this purpose in GRS by, e.g., \cite{Misra:2006qg}. 

For SVD, which is a matrix-based method, we first construct a data matrix $D$ of dimensions $\tau \times M_c$, where $\tau$ is the optimal delay and $M_c$ is the optimal embedding dimension. The delay $\tau$ used here is the same as that used for the CI method. {  For computing $M_c$, although we could have used $D_2$ that are derived as a result of our CI test, this comes with two issues. First, $D_2$ values are unknown for the signals deemed to be S by the CI method. Second, it would make conclusions using SVD implicitly dependent on the conclusions from the CD test, thus defeating the purpose of being another independent test of S/NS behavior. For this reason, we have decided to use an independent method to estimate $M_c$: the False Nearest Neighbour (FNN) technique of the NoLiTSA package \citep{RHODES1997S1149}.} Hence, we define
\begin{equation}D_{ij}=x_{i+(j-1)\tau}.\end{equation}
We perform SVD of $D$ as,
\begin{equation}D=U\Sigma V^{T},\end{equation}
where $U$ and $V$ are the left and right singular vectors, respectively, $\Sigma$ is a diagonal matrix containing the singular values of $D$. It is well known that the temporal dynamics of the timeseries can be understood from the right singular vector $V$, whose first two components are $E1$ and $E2$. The variation of $E1$ with respect to $E2$ gives the dominant temporal dynamics of the timeseries.

In previous works applying SVD on GRS 1915+105 classes \citep{Misra:2006qg}, the distinction between S and NS type dynamics was mostly made by visual inspection. In the present work, we try to take a more quantitative approach. As proposed by \cite{PCA_SVD}, we employ a commonly used topological descriptor, Betti numbers \citep{Betti1870}. Betti number descriptor for a $d-$dimensional manifold is a vector of $d$ integers represented as $\beta=(\beta_0,\beta_1,\cdots,\beta_{d-1})$. In our two-dimensional $E1$ vs $E2$ map, we need to consider only $\beta_0$ (number of connected components or ``blobs") and $\beta_1$ (number of 1-d holes).

No prominent structure is expected to be visible for an S timeseries, and we shall see only a single blob in the plot; $(\beta_0,\beta_1)=(1,0)$. On the contrary, for signals arising from deterministic (possibly C) dynamics, we see a more complex topology with multiple connected components and holes. {  Physically, we want to know whether there are multiple spots where the system's dynamics tends to center around ($\beta_0$), and whether there are any closed orbits that can be identified ($\beta_1$).}

In the ideal case of noise-free test data (such as a Lorenz system), the $E1-E2$ plot becomes very complex with $\beta_i\gg1$; the plots resemble phase portraits. In a general case, the classification between S and NS cases is done based on $\beta_i$ as follows:
\begin{equation*}
    \begin{split}
    \beta_0 + \beta_1 &=1 \Rightarrow \text{S signal},\\
                      &> 1 \Rightarrow \text{NS signal}.
\end{split}
\end{equation*}
To quantitatively estimate these numbers, we use Kernel Density Estimation (KDE) to non-parametrically estimate the distribution function of points on the $E1-E2$ plane. We then identify the over-dense regions by finding the local maxima of this distribution. Simultaneously, we also plot the contour corresponding to the median density (i.e. the density in the $E1-E2$ plane such that 50\% of the points are above this value). 
$\beta_0$ is taken to be the number of such maxima which are found inside the contour. $\beta_1$, i.e. the number of loops, can now be clearly seen from both this density contour as well as the phase portrait.

The results from this method are listed in subsection \ref{subsec: SVD_result}.

\subsection{Principle Component Analysis (PCA) and DBSCAN} \label{subsec: PCA}
PCA is generally used for dimensionality reduction of large data sets with many features per observation.{  A general overview of PCA can be found in \citealt{PCA1987}.} In our application, we utilize an algorithm proposed by \cite{PCA_SVD} (who successfully implemented this to classify the GRS 1915+105 classes), which iteratively computes the eigenvalue ratios of the covariance matrix for different sub-intervals of the timeseries. The implementation details can be found in \cite{PCA_SVD}; we outline the basic procedure here. First, we initialize an empty list $ER_{list}={0,\cdots,0}$ of the same length as the timeseries, whose values are determined using the following algorithm. We split the lightcurve into two equal segments and construct a $2\times 2$ covariance matrix. We calculate the ratio of the eigenvalues ($\lambda_1$ and $\lambda_2$) of this matrix: $r=\lambda_1/\lambda_2$ if $\lambda_1 \ge \lambda_2$. If the data show any dominant direction (as is in an NS timeseries), the larger eigenvalue will be significantly greater than the other, leading to a large $r$. On the contrary, no dominant direction for an S timeseries should exist, and the two eigenvalues should be comparable, hence the ratio is small or even of the order of unity. If the ratio is greater than a predetermined cutoff ($r\ge r_{cut}$), we store $r$ in the list $ER_{list}$ (at the same indices corresponding to the segment whose covariance matrix is being calculated). If the ratio is less than the cutoff value, we do not assign any value to the $ER_{list}$; rather we split each of the two segments further and repeat the process of calculating the eigenvalue ratios for them. This process continues until either the ratio cutoff condition is met or we reach a minimum size of the segments, which we have chosen to be 10 following \citealt{PCA_SVD}. 

We consider three features derived from the $ER$ series (suggested by \cite{PCA_SVD}). They are the following:
\begin{enumerate}
    \item \textbf{MER}: Maximum ER found within the lightcurve. This should be lower for an S signal than in NS cases, indicating a lack of dominant orientation.
    \item \textbf{VAR}: The variance of the ER curve. This is lower in the case of S timeseries since the range of ER is also low for them.
    \item \textbf{Area}: The total area under the ER curve. This is greater for the deterministic or NS cases since the ER stays high for most data.
    
\end{enumerate}
Earlier, such analysis was only done for GRS 1915+105 temporal classes. However, the distinction between S and NS was done manually. No systematic method existed to classify between S and NS classes, which is what we propose in the current work.

We suggest the following method. Each original and denoised signals can be represented as a point in the 3D (MER, VAR, Area) space. The S signals are expected to populate close to the origin due to low MER, VAR, and Area. Identifying this cluster would successfully group all the S signals. The NS signals having orders of magnitude greater values of these variables than S signals are scattered further away from the origin. Since each filtering technique gives different MER, VAR, and Area values, the NS classes are not clustered; they appear as \textit{outliers}. We employ an unsupervised clustering technique -- Density-Based Spatial Clustering of Applications with Noise (DBSCAN), to identify the cluster of S signals and the outliers (i.e., NS).

DBSCAN requires two hyper-parameters to be supplied: (1) the minimum number of points within a cluster and (2) the upper bound on the nearest neighbor distance between two points of a cluster ($\epsilon$). We make $N$ artificial stochastic signals for this classification scheme and determine their PCA parameters. These form a baseline and help us fix the minimum number of points that must be in the cluster for S signals to $N$. Thus, we are left with only the nearest neighbor distance $\epsilon$ as the free parameter. 

\begin{deluxetable*}{cccccccccccccc}
\tabletypesize{\scriptsize}
\tablewidth{\textwidth} 
\tablecaption{Correlation Integral results of IGR J17091-3624\label{tab:CI}}

\tablehead{
\colhead{ObsID} & \colhead{Class}& \colhead{GRS-like}& \colhead{$nmsd$}& \colhead{$D_2$}& \colhead{$nmsd$}& \colhead{$D_2$}& \colhead{$nmsd$}& \colhead{$D_2$}& \colhead{$nmsd$}& \colhead{$D_2$}& \colhead{$nmsd$}& \colhead{$D_2$}& \colhead{Behaviour} \\
\colhead{} & \colhead{}& \colhead{class}& \colhead{(original)}& \colhead{(original)}& \colhead{(ADA)}& \colhead{(ADA)}& \colhead{(NLM)}& \colhead{(NLM)}& \colhead{(GAU)}& \colhead{(GAU)}& \colhead{(BOX)}& \colhead{(BOX)}& \colhead{}
}
\colnumbers
\startdata 
{96420-01-01-00} & {\rom{1}} & {$\chi$}& {1.169} &  {NA} &  {0.862} &  {NA} &  {2.945} &  {NA} &  {0.824} &  {NA} &  {0.793} &  {NA} &  {S}\\
{96420-01-11-00} & {\rom{2}} & {$\phi$}& {1.322} &  {NA} &  {1.807} &  {NA} &  {2.645} &  {NA} &  {1.223} &  {NA} &  {0.815} &  {NA} &  {S}\\
{96420-01-04-01} & {\rom{3}} & {$\nu$}& {0.276} &  {NA} &  {1.035} &  {NA} &  {3.863} &  {8.159} &  {0.837} &  {NA} &  {0.666} &  {NA} &  {S}\\
{96420-01-05-00} & {\rom{4}} & {$\rho$}& {1.300} &  {NA} &  {4.525} &  {5.255} &  {5.861} &  {3.841} &  {4.756} &  {4.340} &  {5.883} &  {4.956} &  {NS}\\
{96420-01-06-03} & {\rom{5}} & {$\mu$}& {1.612} &  {NA} &  {4.332} &  {4.211} &  {8.122} &  {3.522} &  {5.707} &  {3.649} &  {3.764} &  {3.927} &  {NS}\\
{96420-01-09-00} & {\rom{6}} & {$\lambda$}& {0.829} &  {NA} &  {2.418} &  {NA} &  {1.115} &  {NA} &  {2.820} &  {NA} &  {4.311} &  {5.546} &  {S}\\
{96420-01-18-05} & {\rom{7}} & {None}& {0.926} &  {NA} &  {6.055} &  {3.150} &  {4.011} &  {3.069} &  {4.494} &  {3.333} &  {4.811} &  {3.103} &  {NS}\\
{96420-01-19-03} & {\rom{8}} & {None}& {1.482} &  {NA} & {11.814} &  {4.642} &  {17.039} &  {3.958} &  {7.659} &  {4.237} &  {7.149} &  {4.388} &  {NS}\\
{96420-01-35-02} & {\rom{9}} & {$\gamma$}& {0.948} &  {NA} &  {0.457} &  {NA} &  {1.244} &  {NA} &  {1.392} &  {NA} &  {1.151} &  {NA} &  {S}\\
\enddata

\tablecomments{The columns show the following: (1) \textit{RXTE} ObsID; (2) temporal class based on \cite{Altamirano_2011}; (3) corresponding GRS-like temporal class; (4), (6), (8), (10) and (12) are Normalised Mean Square Deviation ($nmsd$) of unfiltered data, ADA, NLM, GAU and BOX respectively; (5), (7), (9), (11) and (13) are the corresponding $D_2$ estimated for each denoising technique (relevant only if $nmsd\ge3$); (14) dynamical behavior of the system.}
\end{deluxetable*}

For choosing the appropriate cutoff value for ER ($r_{cut}$) and the maximum nearest neighbor distance for DBSCAN, we optimize the clustering performance measured using the silhouette score \citep{Rousseeuw1987}. Our work resolves several technical challenges and formalizes employing PCA to find NS behavior in a timeseries. The results of this method have been presented in subsection \ref{subsec: PCA result}.

\section{Results}\label{sec:results}

The central goal of our study is to rigorously examine the x-ray temporal classes in IGR, if indeed it lacks any deterministic components, as reported earlier \citep{Adegoke2020}, including chaotic or fractal dynamics. We seek to investigate whether these temporal classes can be solely attributed to random noise. 
By subjecting data to various denoising techniques, we aim to scrutinize the underlying nature of the observed temporal behaviors in the  IGR lightcurves. These findings hold significant physical implications, which will be discussed later in the paper.

\subsection{CI method results} \label{subsec: CI results}
Table \ref{tab:CI} lists the $nmsd$ statistics for the unfiltered data and the denoised signals. The $D_2$ values are listed for only those cases whose $nmsd \ge 3$ \citep{harikrishnan2006}. We label a temporal class as ``S" or ``NS" if the conclusion of at least three or more filtering techniques agrees. If only two of our four methods agree, then the overall result is inconclusive, and we label it ``S/NS". {  This overall distinction based on all denoising methods is meant to provide only a general idea of how a class behaves with denoising. We see, in most cases, all of our denoising techniques yield the same conclusion.}

\begin{figure*}[t!]
\centering
\includegraphics[width=\textwidth]{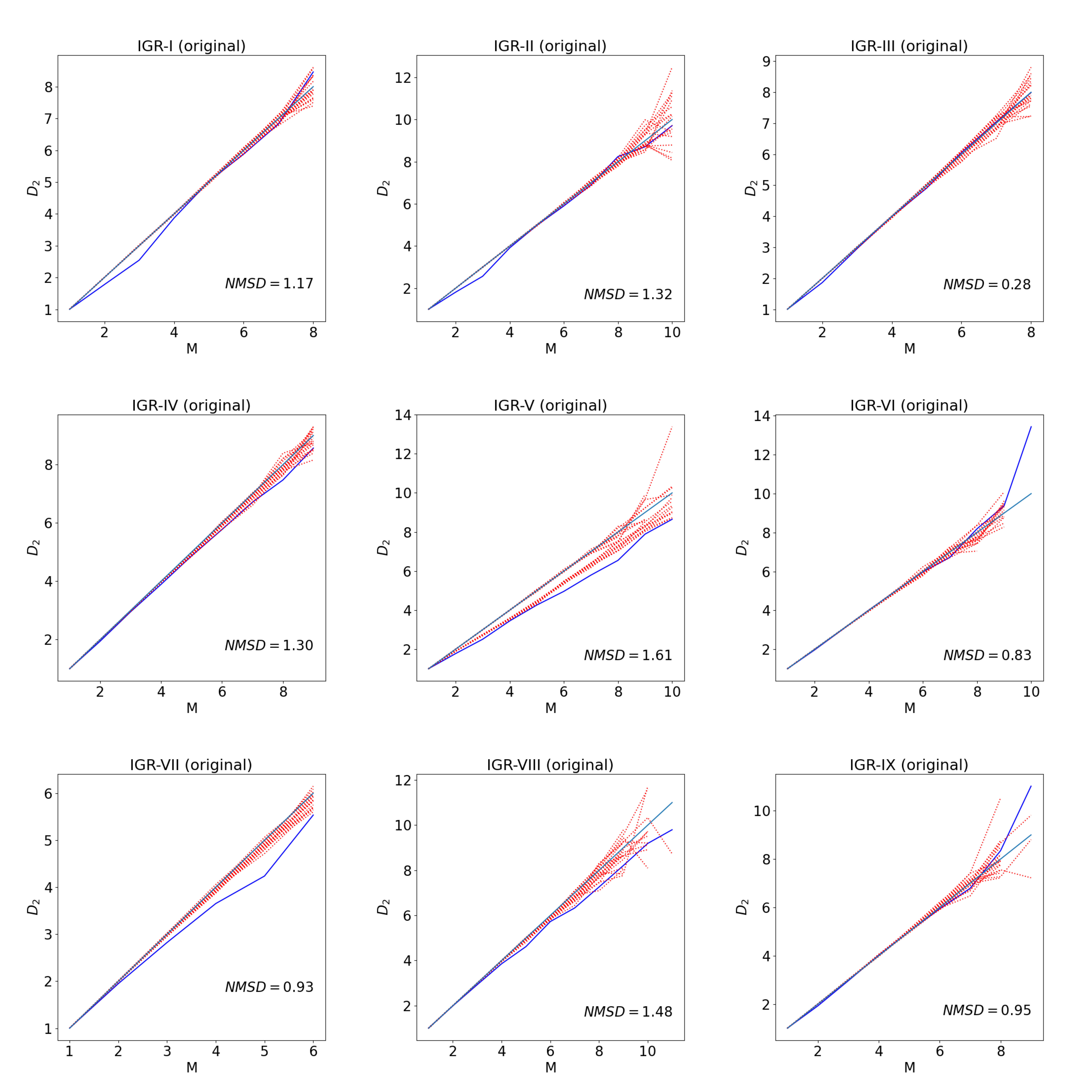}

\caption{$D_2$ vs. $M$ plot for for all classes without any denoising. The red dotted curves are the surrogates, and the dark blue solid curve is the true data. The light blue straight line shows $D_2 = M$, the expected result for an ideal S signal.}
\label{fig: CD ORG}
\end{figure*}

\begin{figure*}[t!]
\centering
\includegraphics[width=\textwidth]{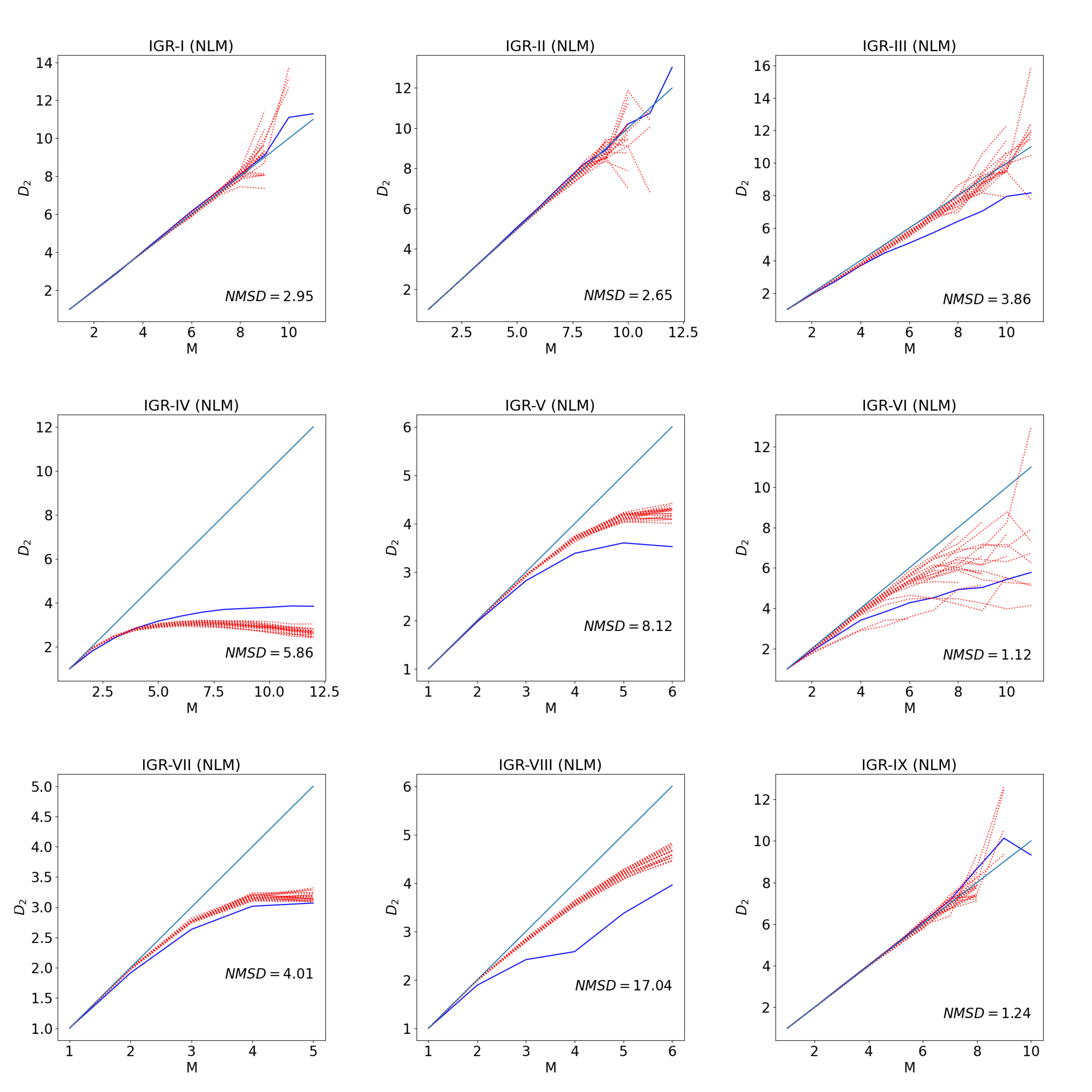}

\caption{$D_2$ vs. $M$ plot for all classes after NLM denoising. The red dotted curves are the surrogates, and the dark blue solid curve is the true data. The light blue straight line shows $D_2 = M$, the expected result for an ideal S signal.}
\label{fig: CD NLM}
\end{figure*}

Figure \ref{fig: CD ORG} shows the $D_2-M$ plots for all temporal classes without any denoising. Here, one can see that in the original timeseries, the true data and the surrogates almost exactly overlap (over the $D_2 = M$ line), thus appearing S, as found earlier by \cite{Adegoke2020}. Figure \ref{fig: CD NLM} shows the $D_2-M$ plots for all temporal classes with NLM denoising. It is evident that upon denoising, $D_2$ of signal from some classes deviates from the $D_2=M$ line, and also, there is a clear distinction of them from that of the corresponding surrogates.

We find that IGR classes \rom{4}, \rom{5}, \rom{7} and \rom{8} show significant deviation from stochasticity. All of them are determined to be NS unanimously by our filtering methods (to avoid repetition, we display results only for NLM denoising).

We also find that regardless of the applied filtering method, classes \rom{1}, \rom{2}, and \rom{9} always turn out to be S. Additionally, the classes \rom{3} and \rom{6} are found to be S for all denoising techniques except NLM (for \rom{3}) and GAU (for \rom{6}). Therefore, we can not rule out the null hypothesis that these IGR classes truly originate from stochastic physical processes. 

This is important as it confirms that the observed signs of determinism are not artifacts from the denoising procedure. Our methodology is capable of capturing the lightcurve's inherent stochasticity. 

For all of these classes, the $D_2$ values obtained from each method are quite close to one another (within $\pm 1$). This agreement among all the denoising techniques validates our approach and strongly indicates the presence of some low-dimensional physical processes in these four temporal classes. It is interesting to note that the corresponding GRS-like classes of IGR-\rom{1}, \rom{2}, and \rom{9} are $\chi$, $\phi$, and $\gamma$ respectively; these GRS-classes are well known to be S (see \citealt{Adegoke2018}, \citealt{Misra:2006qg}). Similarly, the GRS-like classes of \rom{4} and \rom{5} are $\rho$ and $\mu$ respectively, which are known to be NS states of GRS ($\rho$ is a limit cycle and $\mu$ is fractal). Hence, CI results hint at the close correspondence between GRS's and  IGR's temporal classes.

\subsection{Results from SVD} \label{subsec: SVD_result}
\begin{figure*}[p!]
\centering

\includegraphics[width=\textwidth]{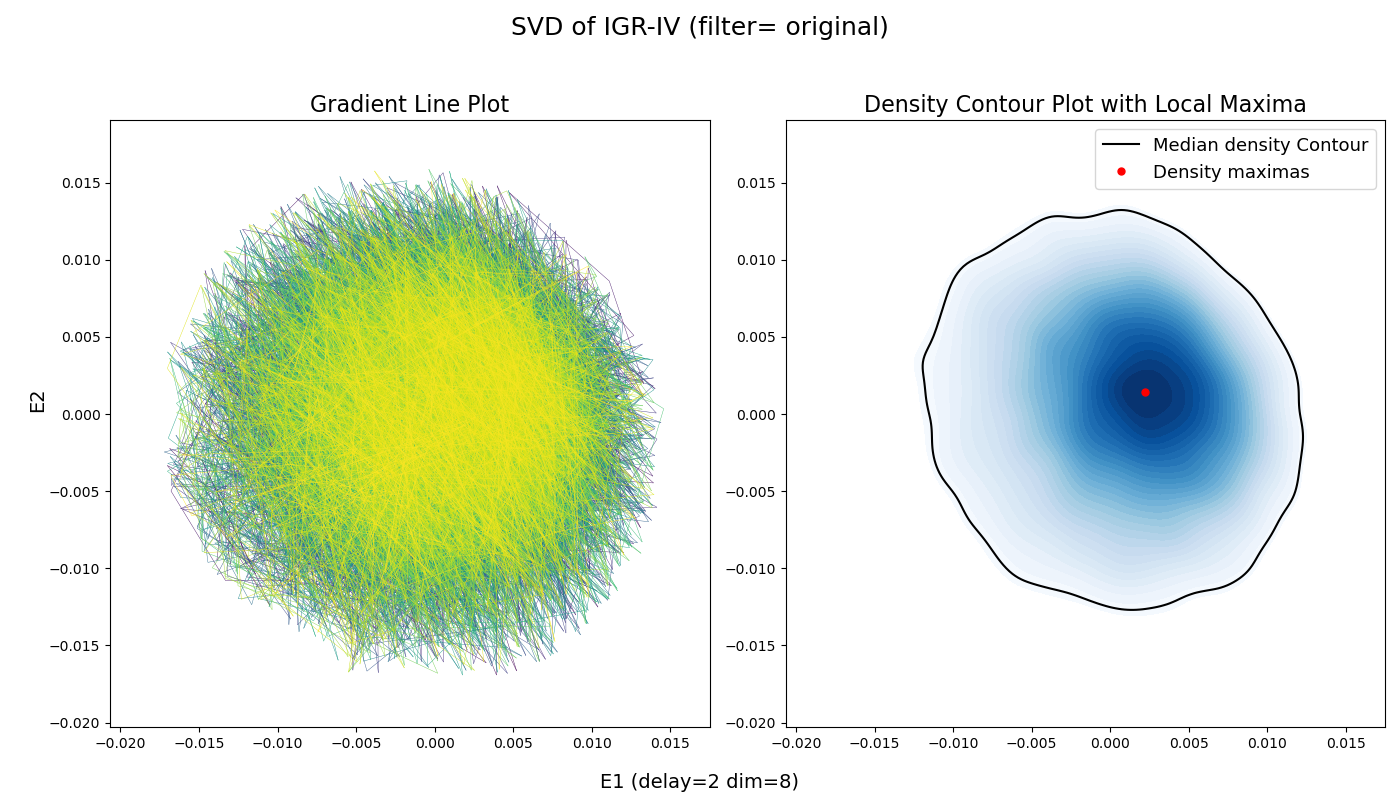}
\includegraphics[width=\textwidth]{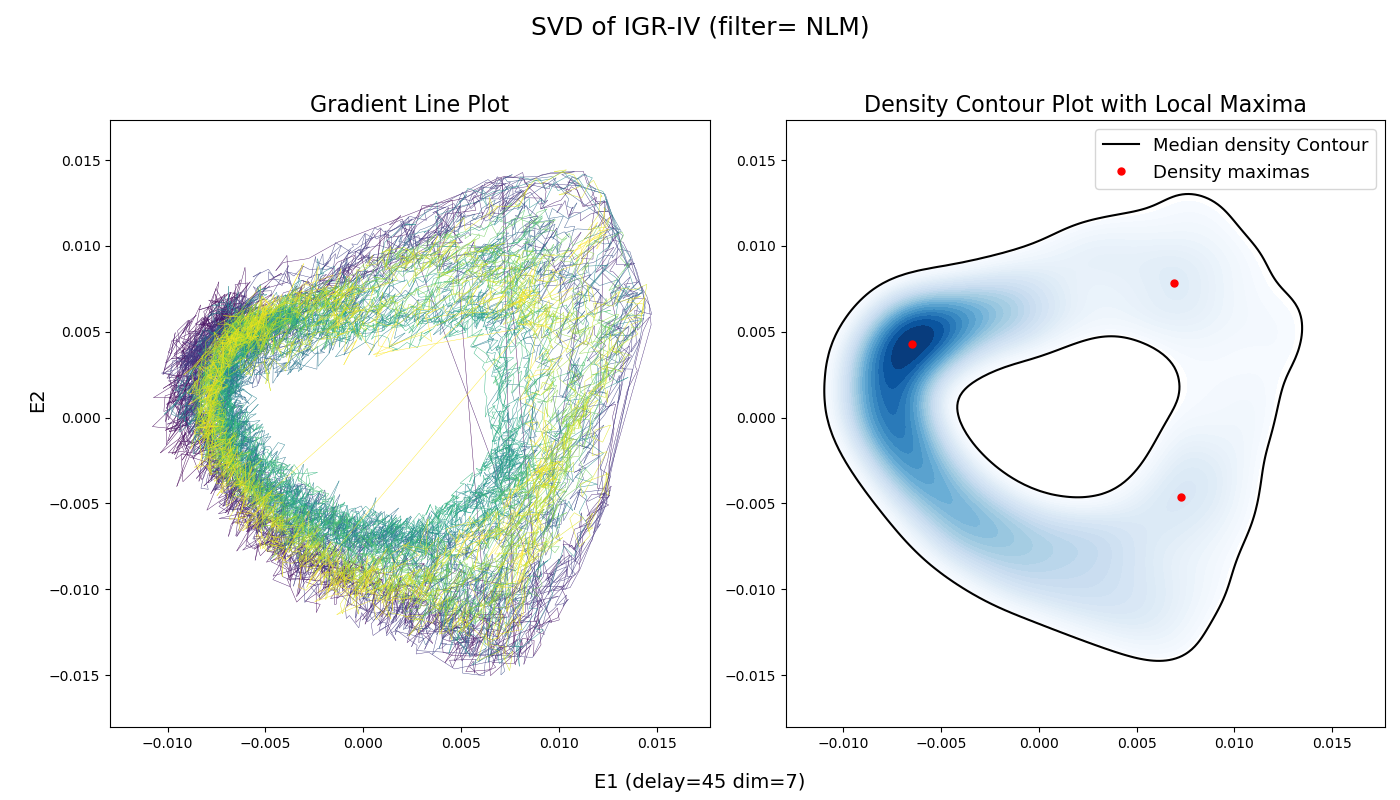}

\caption{SVD plots of IGR-\rom{4} original data, and after NLM denoising. The optimal embedding dimensions and delay are mentioned for each of them.}
\label{fig: SVD IGR-4}
\end{figure*}

\begin{figure*}[p!]
\centering

\includegraphics[width=\textwidth]{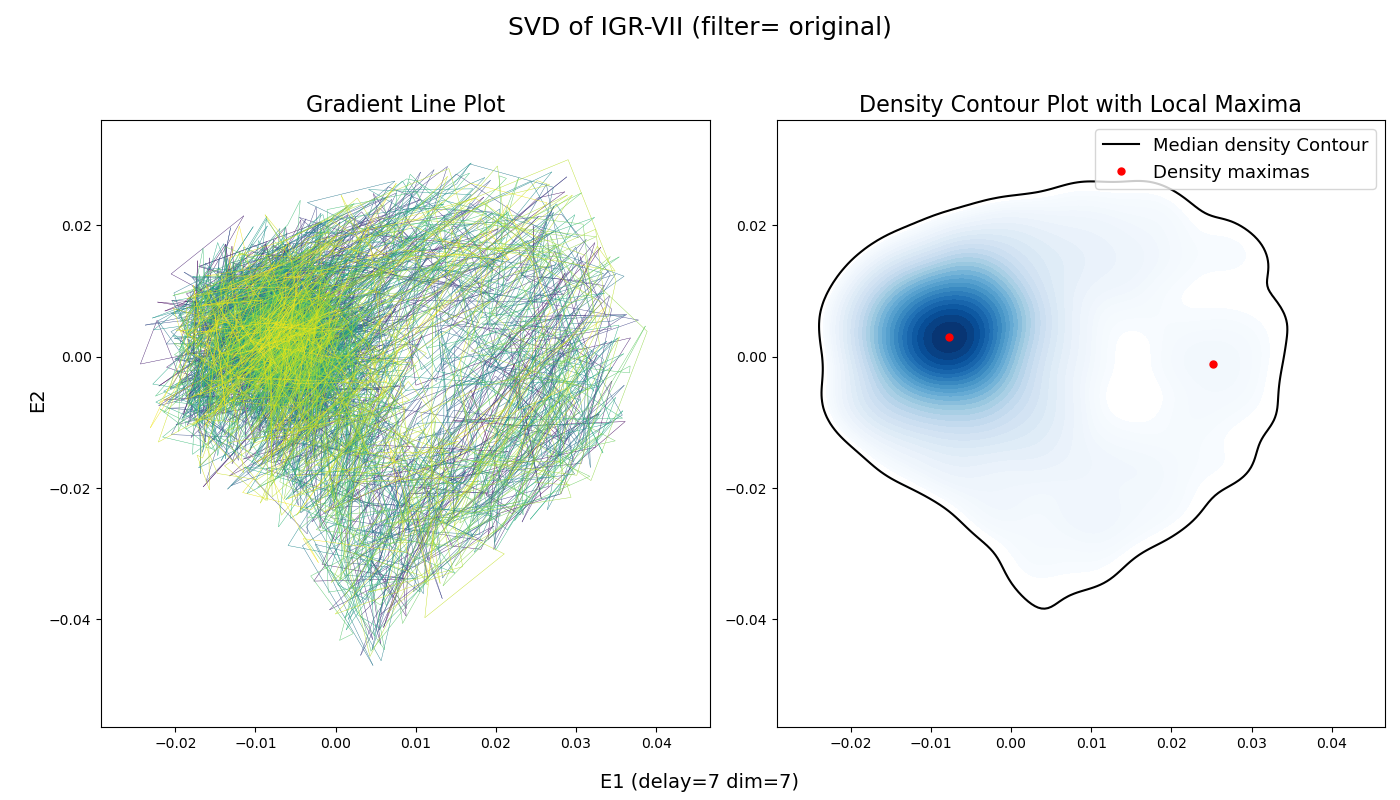}
\includegraphics[width=\textwidth]{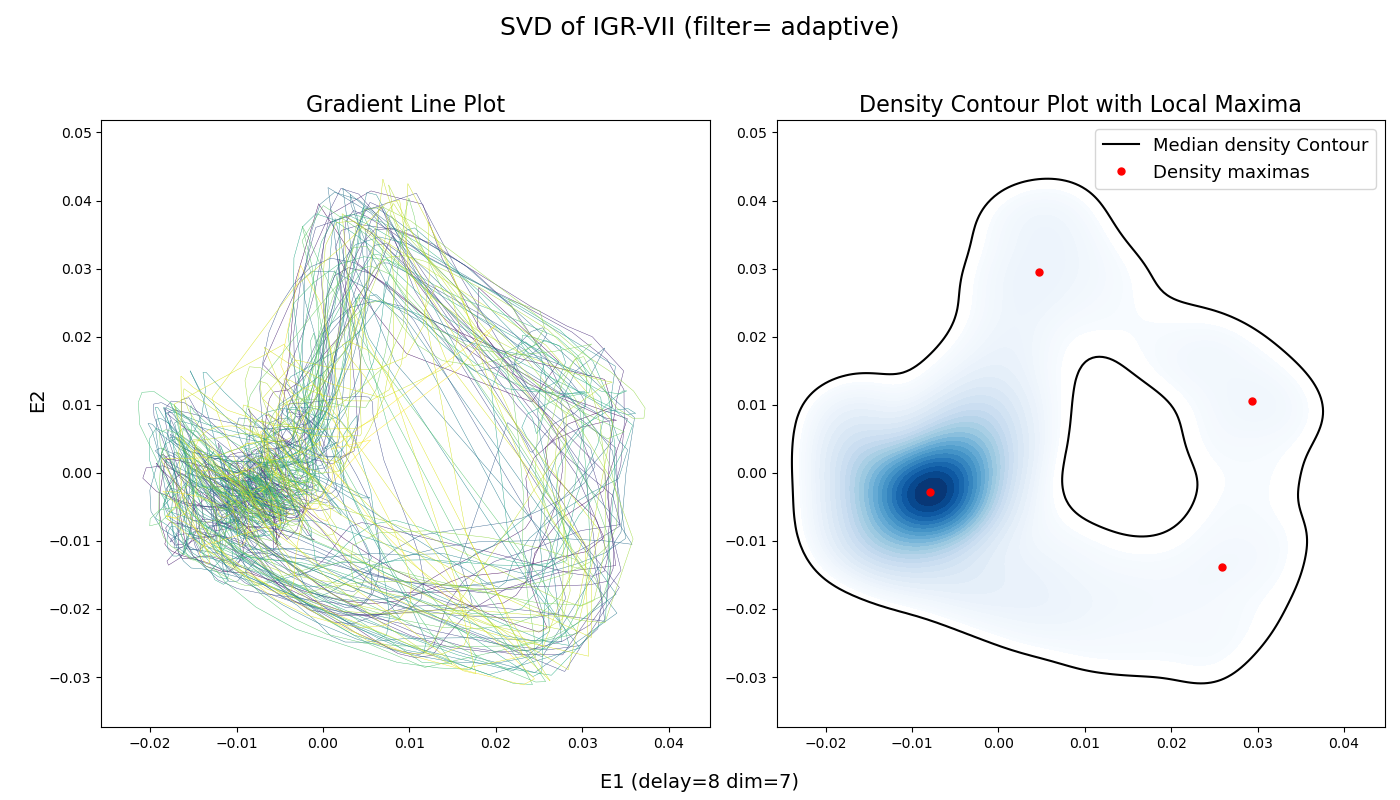}
\caption{SVD plots of IGR-\rom{7} original data, and after ADA denoising. The optimal embedding dimensions and delay are mentioned for each of them.}
\label{fig: SVD IGR-7}
\end{figure*}

\begin{figure*}[p!]
\centering

\includegraphics[width=\textwidth]{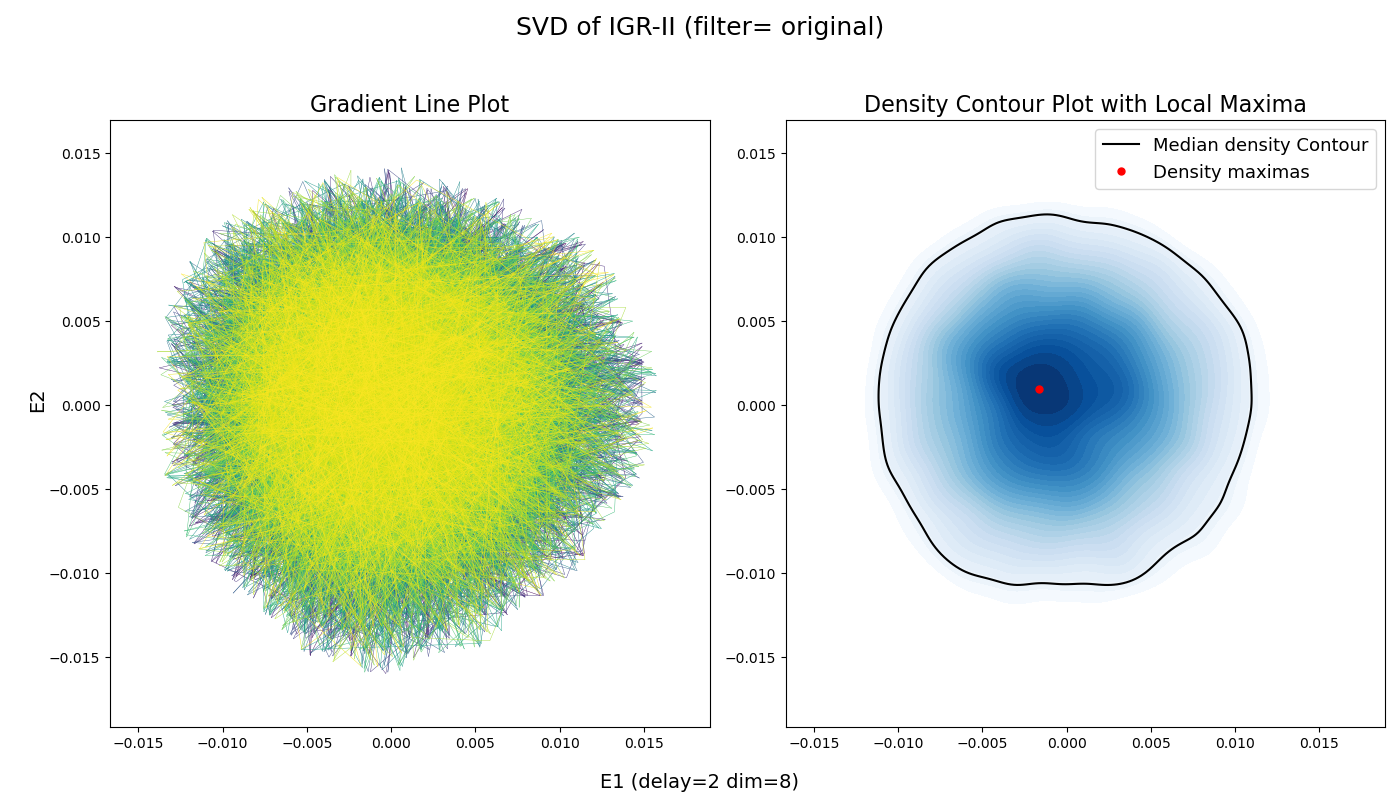}
\includegraphics[width=\textwidth]{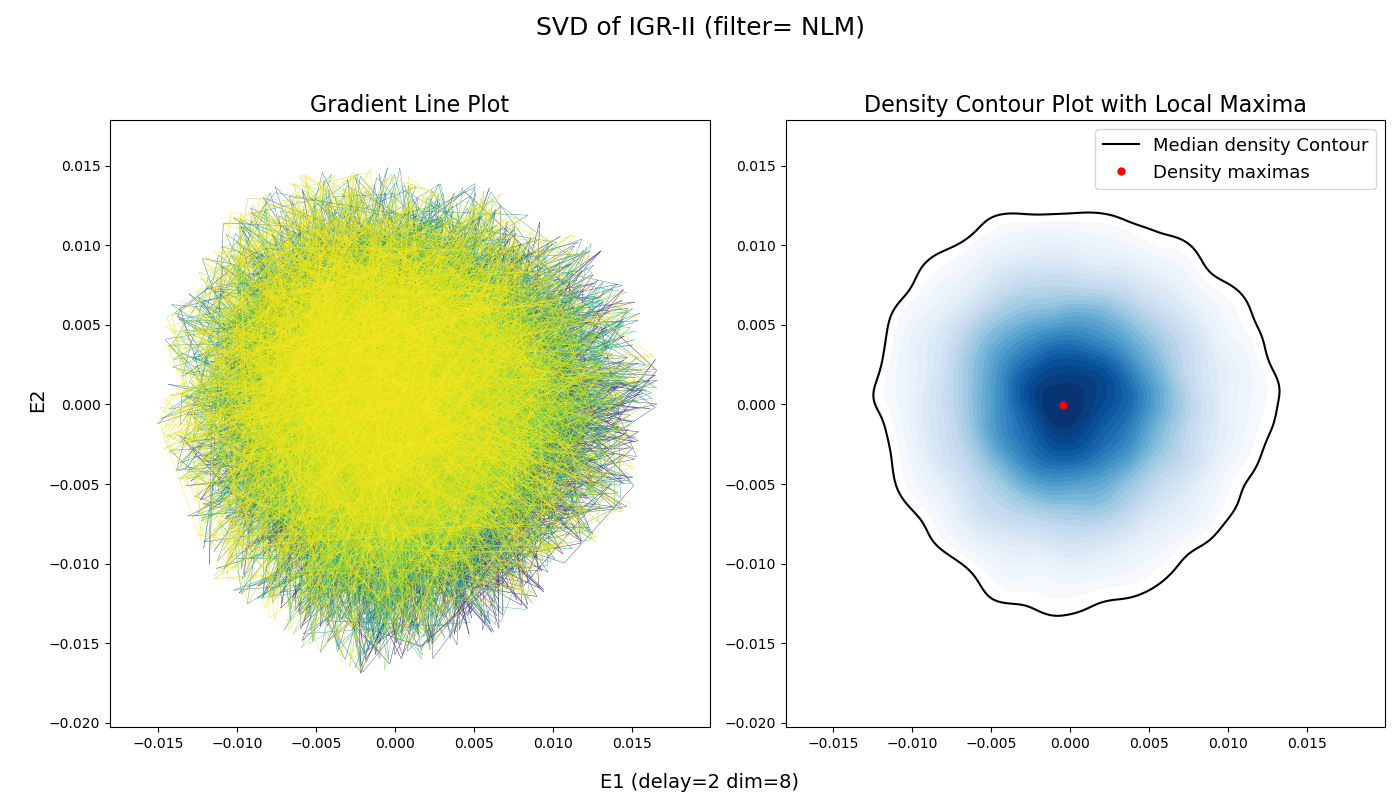}
\caption{SVD plots of IGR-\rom{2} original data, and after ADA denoising. The optimal embedding dimensions and delay are mentioned for each of them.}
\label{fig: SVD IGR-2}
\end{figure*}

 Through SVD, we find that classes \rom{5}, \rom{6}, \rom{7}, and \rom{8} show complex trajectories with multiple loops (high $\beta_1$) and dense regions, reminiscent of attractor behavior (corresponding to high $\beta_0$). While the other classes (\rom{1}, \rom{2}, \rom{3}, and \rom{9}) show no topological structures like rings or multiple connected components. They have the structure typically expected of S signals, i.e., $(\beta_0, \beta_1) = (1,0)$. The classs \rom{4} shows complex dynamics via NLM and GAU denoising techniques. Especially, with NLM denoising, we find almost textbook attractor nature (see Figure \ref{fig: SVD IGR-4}). In our figures, the left panel shows the evolution of the system in the embedded space, and the right panel shows the estimated distribution over the $E1-E2$ space and the presence of loops and multiple over-dense regions can be seen from it.  To highlight the gain in clarity achieved by our denoising techniques, we show the SVD portraits of IGR-\rom{7} in Figure \ref{fig: SVD IGR-7}. In contrast, Figure \ref{fig: SVD IGR-2} demonstrates the single blob-like nature of a typical S timeseries (IGR-\rom{1}). We also find that the original time series for classes \rom{6}, \rom{7}, and  \rom{8} exhibits complex dynamics \textit{without} any denoising. As previously the IGR lightcurves were not tested using SVD, this is a noteworthy observation.
 
 Unlike the other testing methods we have used for this study, the SVD phase portraits are not limited to only a binary distinction between S and NS. This method provides a more tangible perspective of how the dynamics of a potentially deterministic source work. The topological parameters, such as the Betti numbers, can be connected to the projection of phase space features such as attractors, limit cycles, etc. 

Table \ref{tab:SVD} summarizes the observed dynamical nature based on the SVD of each temporal class subjected to the denoising methods.

\begin{deluxetable*}{ccccccccc}
\tabletypesize{\scriptsize}
\tablewidth{\textwidth} 
\tablecaption{Singular Value Decomposition results of IGR J17091-3624\label{tab:SVD}}

\tablehead{
\colhead{ObsID} & \colhead{Class}& \colhead{GRS-like class}& \colhead{original}& \colhead{ADA}& \colhead{NLM}& \colhead{GAU}& \colhead{BOX}& \colhead{Overall}
}
\colnumbers
\startdata 
{96420-01-01-00} & {\rom{1}} & {$\chi$} & {S} & {S} & {S} & {S} & {S} & {S}\\
{96420-01-11-00} & {\rom{2}} & {$\phi$} & {S} & {S} & {S} & {S} & {S} & {S}\\
{96420-01-04-01} & {\rom{3}} & {$\nu$} & {S} & {S} & {S} & {NS} & {S} & {S}\\
{96420-01-05-00} & {\rom{4}} & {$\rho$} & {S} & {s} & {NS} & {NS} & {S} & {S/NS}\\
{96420-01-06-03} & {\rom{5}} & {$\mu$} & {S} & {NS} & {NS} & {NS} & {NS} & {NS}\\
{96420-01-09-00} & {\rom{6}} & {$\lambda$} & {NS} & {NS} & {S} & {NS} & {NS} & {NS}\\
{96420-01-18-05} & {\rom{7}} & {None} & {NS} & {NS} & {NS} & {NS} & {NS} & {NS}\\
{96420-01-19-03} & {\rom{8}} & {None} & {NS} & {NS} & {NS} & {NS} & {NS} & {NS}\\
{96420-01-35-02} & {\rom{9}} & {$\gamma$} & {S} & {S} & {S} & {S} & {S} & {S}\\
\enddata

\tablecomments{The columns show the following: (1) \textit{RXTE} ObsID; (2) temporal class based on \cite{Altamirano_2011}; (3) corresponding GRS-like class; (4), (5), (6), (7), and (8) are the results from SVD plots with no filters, ADA, NLM, GAU and BOX respectively; (9) is the overall conclusion from all methods.}
\end{deluxetable*}

\begin{deluxetable*}{ccccccccc}
\tabletypesize{\scriptsize}
\tablewidth{\textwidth} 
\tablecaption{Principle Component Analysis results of IGR J17091-3624\label{tab:PCA}}

\tablehead{
\colhead{ObsID} & \colhead{Class}& \colhead{GRS-like class}& \colhead{original}& \colhead{ADA}& \colhead{NLM}& \colhead{GAU}& \colhead{BOX}& \colhead{Overall}
}
\colnumbers
\startdata 
{96420-01-01-00} & {\rom{1}} & {$\chi$}& {S} & {S} & {S} & {S} & {S} & {S}\\
{96420-01-11-00} & {\rom{2}} & {$\phi$}& {S} & {S} & {S} & {S} & {S} & {S}\\
{96420-01-04-01} & {\rom{3}} & {$\nu$}& {S} & {S} & {S} & {S} & {S} & {S}\\
{96420-01-05-00} & {\rom{4}} & {$\rho$}& {S} & {S} & {NS} & {NS} & {NS} & {NS}\\
{96420-01-06-03} & {\rom{5}} & {$\mu$}& {S} & {NS} & {NS} & {NS} & {NS} & {NS}\\
{96420-01-09-00} & {\rom{6}} & {$\lambda$}& {S} & {S} & {NS} & {S} & {NS} & {S/NS}\\
{96420-01-18-05} & {\rom{7}} & {None}& {S} & {NS} & {NS} & {NS} & {NS} & {NS}\\
{96420-01-19-03} & {\rom{8}} & {None}& {S} & {S} & {NS} & {NS} & {NS} & {NS}\\
{96420-01-35-02} & {\rom{9}} & {$\gamma$}& {S} & {S} & {S} & {S} & {S} & {S}\\
\enddata

\tablecomments{The columns show the following: (1) \textit{RXTE} ObsID; (2) temporal class based on \cite{Altamirano_2011}; (3) corresponding GRS-like class; (4), (5), (6), (7), and (8) are the results from PCA with no filters, ADA, NLM, GAU and BOX respectively; (9) is the overall conclusion from all methods.}
\end{deluxetable*}

\subsection{PCA and DBSCAN results} \label{subsec: PCA result}

\begin{figure}[h]    
    \centering
    \includegraphics[width=\linewidth, trim={0cm 0cm 0cm 2.3cm},clip]{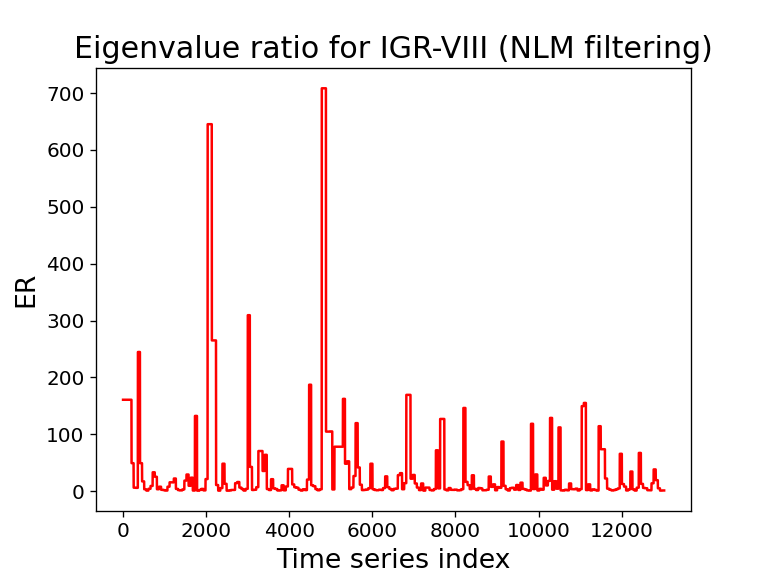}
    \caption{An example of a typical ER curve (IGR \rom{8} post NLM filtering).}
\label{fig: ER curve}
\end{figure}

\begin{figure}[h]    
    \centering
    \includegraphics[width=\linewidth, trim={0cm 1cm 2cm 2cm},clip]{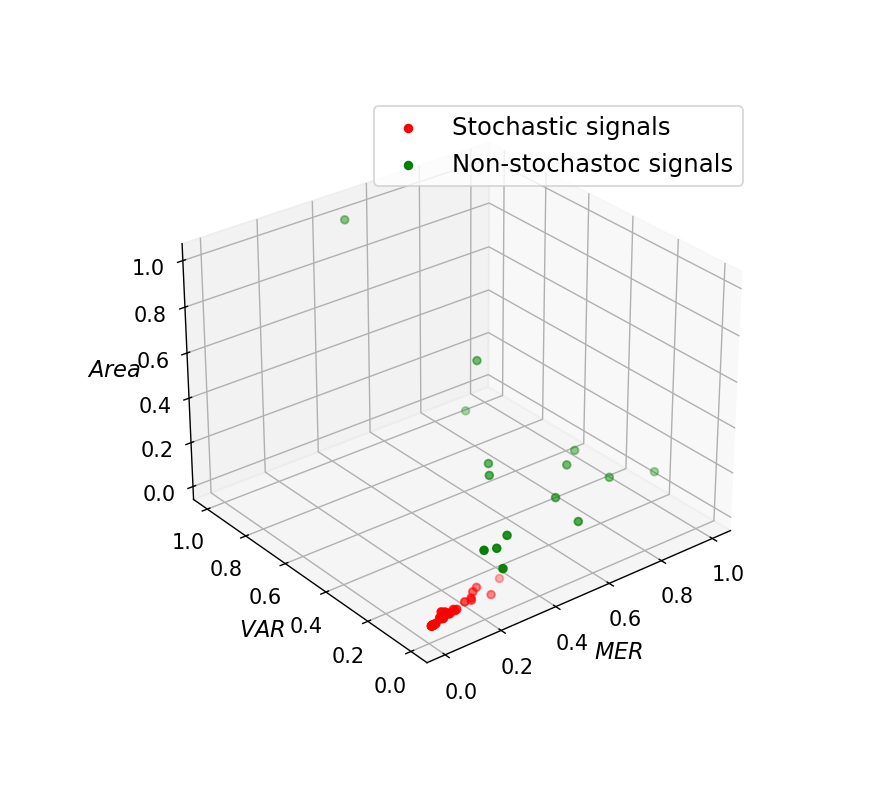}
    \caption{Scatter plot of the PCA parameters for all signals (filtered and unfiltered). Red points are the S timeseries, while green points are NS.}
\label{fig: classification}
\end{figure}

As outlined in subsection \ref{subsec: PCA}, the PCA algorithm has been applied to the original and filtered timeseries sets, and the following results have been obtained. Figure \ref{fig: ER curve} shows the ER curve of a typical NS signal (IGR-\rom{8}, NLM filtered) for demonstration. We find that the optimum eigenvalue ratio cutoff is $124$, and the $\epsilon$ parameter of DBSCAN is $40$ which maximizes the value of clustering silhouette score. Our score is $0.693$, which is acceptable\footnote{Silhouette score $\in (-1, 1)$. The closer it is to $1$, the more efficient the cluster configuration.}. Figure \ref{fig: classification} shows the actual scatter of the cluster and the outliers. The green points are the outliers, which we identify as the NS timeseries.

This test yields similar conclusions to CI and surrogate analysis test. Class \rom{5} and \rom{7} as found to be consistently NS by all denoising methods, while \rom{1}, \rom{3}. and \rom{9} are always S. NLM and BOX yields identical results.

\section{Summary and conclusions}\label{sec:Discussion}

We have explored the timeseries analysis of the observed X-ray
binary IGR J17091-3624 to understand the underlying nonlinear properties. One of the hindrances against this analysis is the noise contamination of the data. We have addressed the issue of Poisson noise contamination in the available X-ray lightcurves of IGR J17091-3624 (nine temporal classes) by applying four different filtering techniques: ADA (adaptive), NLM (non-local means), GAU (Gaussian convolution), and BOX (Boxcar convolution or moving average). To check for NS (or even C) behavior in the filtered signals, we have applied a relatively new method in the present context like the PCA, along with more traditional ones, such as CI and SVD. The PCA and SVD tests have been improved upon by making them more objective. Our findings are summarized in Table \ref{tab:summary}.

\begin{deluxetable}{ccccccc}
\tabletypesize{\scriptsize}
\tablewidth{\textwidth} 
\tablecaption{Summary of S/NS behavior \label{tab:summary}}

\tablehead{
\colhead{ObsID}& \colhead{Class}& \colhead{GRS-like class}& \colhead{CI}& \colhead{SVD}& \colhead{PCA}& \colhead{Overall}
}

\colnumbers
\startdata 
{96420-01-01-00} & {\rom{1}} & {$\chi$}& {S}     & {S}     & {S}  & {S}\\ 
{96420-01-11-00} & {\rom{2}} & {$\phi$}& {S}     & {S}     & {S}  & {S}\\
{96420-01-04-01} & {\rom{3}} & {$\nu$}& {S}     & {S}     & {S}  & {S}\\
{96420-01-05-00} & {\rom{4}} & {$\rho$}& {NS}    & {S/NS}  & {NS} & {NS}\\
{96420-01-06-03} & {\rom{5}} & {$\mu$}& {NS}    & {NS}    & {NS} & {NS}\\
{96420-01-09-00} & {\rom{6}} & {$\lambda$}& {S}     & {NS}    & {S}  & {S/NS}\\
{96420-01-18-05} & {\rom{7}} & {None}& {NS}    & {NS}    & {NS} & {NS}\\
{96420-01-19-03} & {\rom{8}} & {None}& {NS}    & {NS}    & {NS} & {NS}\\
{96420-01-35-02} & {\rom{9}} & {$\gamma$}& {S}     & {S}     & {S}  & {S}\\
\enddata

\tablecomments{The columns show the following: (1) \textit{RXTE} ObsID; (2) temporal class based on \cite{Altamirano_2011}; (3) corresponding GRS-like class; (4), (5), and (6) are `S/NS' classification result using CI, SVD, and PCA respectively; (7) Our final conclusion regarding the behavior of a temporal class when 3 or more methods agree.}
\end{deluxetable}

\begin{deluxetable*}{cccccccccccc}
\tabletypesize{\scriptsize}
\tablewidth{\textwidth}
\tablecaption{Compilation of spectral nature of IGR J17091–3624 classes from \citep{Adegoke2020} with our S/NS classification\label{tab:spec+S/NS}}
\label{spectim}
\tablehead{
\colhead{ObsID} & \colhead{Class} & \colhead{GRS-like class}& \colhead{Behaviour} & \colhead{\textit{diskbb}} & \colhead{PL} & \colhead{GA} & \colhead{SI} & \colhead{$\chi^2$/dof} & \colhead{State} & \colhead{$T_{in}$} & \colhead{$F\times10^{-10}$}
}
\colnumbers
\startdata
96420-01-01-00 & \rom{1} & {$\chi$}& S & 10.0 & 89.1 & 0.9 & $2.26^{+0.02}_{-0.02}$ & 1.12 & PD & $1.11^{+0.05}_{-0.05}$ & 11.949 \\
96420-01-11-00 & \rom{2} & {$\phi$}& S & 18.5 & 81.5 & -- & $2.33^{+0.04}_{-0.04}$ & 1.04 & PD & $1.13^{+0.02}_{-0.02}$ & 6.238 \\
96420-01-04-01 & \rom{3} & {$\nu$}& S & 38.0 & 62.0 & -- & $2.25^{+0.07}_{-0.07}$ & 0.82 & PD & $1.20^{+0.01}_{-0.01}$ & 9.730 \\
96420-01-05-00 & \rom{4} & {$\rho$}& S/NS & 41.6 & 58.0 & 0.4 & $2.34^{+0.05}_{-0.05}$ & 1.06 & PD & $1.20^{+0.01}_{-0.01}$ & 9.676 \\
96420-01-06-03 & \rom{5} & {$\mu$}& NS & 54.4 & 45.6 & -- & $2.55^{+0.07}_{-0.07}$ & 1.28 & D-P & $1.43^{+0.01}_{-0.01}$ & 9.313 \\
96420-01-09-00 & \rom{6} & {$\lambda$}& S/NS & 65.8 & 34.2 & -- & $2.55^{+0.09}_{-0.09}$ & 1.37 & DD & $1.85^{+0.03}_{-0.03}$ & 10.099 \\
96420-01-18-05 & \rom{7} & {None}& NS & 63.4 & 36.6 & -- & $2.61^{+0.15}_{-0.15}$ & 0.65 & DD & $1.50^{+0.03}_{-0.03}$ & 8.603 \\
96420-01-19-03 & \rom{8} & {None}& NS & 73.4 & 26.4 & 0.2 & $2.29^{+0.12}_{-0.12}$ & 1.30 & DD & $1.78^{+0.02}_{-0.02}$ & 10.596 \\
12406          & \rom{9} & {$\gamma$}& S & 50.5 & 49.5 & -- & $0.58^{+0.80}_{-0.80}$ & 1.016 & PD & $1.53^{+0.15}_{-0.15}$ & 1.220 \\
\enddata
\tablecomments{Columns: (1) \textit{RXTE} ObsID except for IGR-\rom{9}, where \textit{Chandra} observation has been used; (2) temporal class based on \cite{Altamirano_2011}; (3) corresponding GRS-like class; (4) The system's behavior as found in our study; (5) \% of multi-color blackbody component; (6) \% of powerlaw component; (7) \% Gaussian line component (XSPEC model gauss); (8) powerlaw photon spectral index; (9) reduced $\chi^2$; (10) spectral state (DD: disk dominated; D-P: disk-powerlaw contributed; PD: powerlaw dominated); (11) \textit{diskbb} temperature in units of keV; (12) Total flux in the energy range 3-25 keV in units of $erg\,cm^{-2}\,s^{-1}$. For the \textit{Chandra} the energy range is 0.5-8 keV.}
\end{deluxetable*}

IGR classes \rom{1}, \rom{2}, \rom{3}, and \rom{9} are unanimously found to be S by all our classification methods. Similarly, classes IGR-\rom{5}, IGR-\rom{7}, and IGR-\rom{8} have been unanimously found to be NS. For IGR-\rom{4}, although it turns out to be NS by CI and PCA, SVD analysis is inconclusive. We still denote it to be NS since SVD with NLM (one of our best-suited methods) has shown very clear signs of determinism. In the case of IGR-\rom{6}, it is difficult to come to a definite conclusion within the scope of our chosen filtering and testing techniques. It is interesting to note that \cite{Adegoke2020} suspected IGR-\rom{5} and IGR-\rom{8} to be possibly NS by increasing the binning of the timeseries to 0.5s, which we confirm by more rigorous analysis. 

For IGR, we consider the spectral analysis by \cite{Adegoke2020} and combine them with the present temporal/timeseries analysis, as reported in Table \ref{spectim}. Following \cite{Adegoke2018}, the classes identified with S and powerlaw-dominated (PD) states are understood to be General Advective Accretion Flow (GAAF, \citealt{Rajesh&Mukhopadhyay2010}), i.e., the classes IGR-\rom{1}, \rom{2}, \rom{3}, \rom{9}. The classes identified with NS and disk-dominated (DD) states correspond to the Keplerian accretion disk \citep{Shakura-Sunyaev_1973}, i.e., IGR-\rom{7}, \rom{8}. The class IGR-\rom{4} appears to be in transition between Advection Dominated Accretion Flow (ADAF, \citealt{Narayan&Yi1994}) and GAAF, possibly during changes in accretion rate. IGR-\rom{6} shows characteristics intermediate between Keplerian and slim disks \citep{Abramowics1988}, while IGR-\rom{5} suggests a significant GAAF component within an otherwise Keplerian disk, exhibiting NS temporal behavior.
This analysis confirms that IGR J17091-3624, like GRS 1915+105, transitions between various accretion modes. Earlier noise-contaminated analyses suggested only GAAF and slim disk modes, creating an apparent discontinuity in the evolution of accretion states. Our present work reveals the complete picture, showing how a slim disk can evolve through the Keplerian disk phase before converting to a GAAF with decreasing accretion rate. For detailed classifications of accretion flows based on combined timeseries and spectral behaviors, see \citealt{Adegoke2018}.

One key outcome of our extensive analysis of IGR J17091–3624 dynamical properties is the following: its apparent stochastic behavior found in the previous study by \cite{Adegoke2020} is an artifact of Poisson noise domination. By using contemporary denoising techniques, we have discovered that  IGR J17091–3624 has the potential to host a complex underlying dynamical system similar to GRS 1915+105. This strengthens the previous arguments in the literature for these two sources being twins. 

Regarding our denoising techniques, NLM and ADA satisfy all the conditions that we had laid out at the beginning of section \ref{sec:proposed filtering}. ADA had been used previously in the context of non-linear timeseries analysis \citep{Tung2011}, and we have utilized it in the context of astrophysical data. However, the application of NLM in the context of non-linear timeseries has been a novelty in our work, especially in the astrophysical context. We have also found that very simple techniques like moving averages and Gaussian convolution mostly agree with the more advanced methods and might be used as a ``litmus" test before proceeding to them. This has been the case in our study, where we started with the simpler methods and progressed to the more rigorous ones. We expect that the methods presented in this work will be useful in analyzing other noise-contaminated timeseries of astrophysical origin.

\section*{Acknowledgements} \label{sec:ack}

A.G. acknowledges the financial support from KVPY, DST, India. {  The authors thank the referee for carefully reading the manuscript with detailed comments and suggestions, which helped to improve the presentation and understand a certain part of the work better. Thanks are also due to Neelam Sinha of CBR, Bangalore, for the discussion, particularly related to Autoencoder-based analysis, which we finally removed from the manuscript. }

\bibliography{sample631}{}
\bibliographystyle{aasjournal}

\end{document}